\documentclass[]{emulateapj}
\usepackage{graphicx}
\usepackage{color}
\usepackage{bm}
\usepackage{amssymb}
\usepackage{amsmath}
\usepackage{mdwlist}
\usepackage{natbib}
\usepackage{subfigure}
\usepackage{booktabs}
\usepackage[flushleft]{threeparttable}
\newcommand{\icarus}{Icarus}  
\newcommand{\pa}{\partial}
\newcommand{\mb}{\boldsymbol}

\newcommand{\bgeq}{\begin{equation}}
\newcommand{\edeq}{\end{equation}}

\begin{document} 
\title{Pebble Accretion in Turbulent Protoplanetary Disks}
\shorttitle{Pebble Accretion in MRI Turbulence}

\author{Ziyan Xu\altaffilmark{1,2}, Xue-Ning Bai\altaffilmark{3,4,5} \& Ruth A. Murray-Clay\altaffilmark{6}}
\shortauthors{Xu, Bai \& Murray-Clay}

\altaffiltext{1}{Kavli Institute for Astronomy and Astrophysics, Peking University, Beijing 100871, China; ziyanx@pku.edu.cn}
\altaffiltext{2}{Department of Astronomy, Peking University, Beijing 100871, China}
\altaffiltext{3}{Institute for Advanced Study, Tsinghua University, Beijing 100084, China;
xueningbai@gmail.com}
\altaffiltext{4}{Tsinghua Center for Astrophysics, Tsinghua University, Beijing 100084, China}
\altaffiltext{5}{Institute for Theory and Computation, Harvard-Smithsonian Center for Astrophysics, 60 Garden St., MS-51, Cambridge, MA 02138}
\altaffiltext{6}{Department of Astronomy and Astrophysics, University of California, Santa Cruz, CA 95064}

\begin{abstract}
It has been realized in recent years that the accretion of pebble-sized dust particles onto planetary cores is an
important mode of core growth, which enables the formation of giant planets at large distances and assists planet
formation in general. The pebble accretion theory is built upon the orbit theory of dust particles in a laminar protoplanetary disk (PPD). For sufficiently large core mass (in the ``Hill regime"), essentially all particles of appropriate sizes entering the Hill sphere can be captured. However, the outer regions of PPDs are expected to be weakly turbulent due to the magnetorotational instability (MRI), where turbulent stirring of particle orbits may affect the efficiency of pebble accretion. We conduct shearing-box simulations of pebble accretion with different levels of MRI turbulence (strongly turbulent assuming ideal magnetohydrodynamics, weakly turbulent in the presence of ambipolar diffusion, and laminar) and different core masses to test the efficiency of pebble accretion at a microphysical level. We find that accretion remains efficient for marginally coupled particles (dimensionless stopping time $\tau_s\sim0.1-1$) even in the presence of strong MRI turbulence. Though more dust particles are brought toward the core by the turbulence, this effect is largely canceled by a reduction in accretion probability. As a result, the overall effect of turbulence on the accretion rate is mainly reflected in the changes in the thickness of the dust layer.  On the other hand, we find that the efficiency of pebble accretion for strongly coupled particles (down to $\tau_s\sim0.01$) can be modestly reduced by strong turbulence for low-mass cores.
\end{abstract}


\section{Introduction} 

In the past decade, direct imaging surveys have discovered a number of giant planets at wide separations
from their host stars (see \citealp{Bowler16} for an up-to-date review). Examples include the HR 8799 system with
four giant planets in the gas-giant range at separations of 14-70 au \citep{HR8799}, and the possible gas giant Fomalhaut b
\citep{Kalas2008} located at 119 au from its central star. Although the overall occurrence rate of such widely
separated gas giants is relatively low ($\sim1\%$ for separation $\gtrsim10$ au, \citealp{Bowler16}), the fact that such planets
exist already poses challenges to the conventional theory of planet formation.

Planet formation takes place in the gaseous and dusty protoplanetary disks (PPDs) surrounding young
protostars. In the standard core-accretion theory for the formation of giant planets, a crucial step is the
formation of a sufficiently massive solid core to enable runaway gas accretion from its parent PPD
\citep{mizuno1980,coreacc}. The critical core mass depends on the gas opacity, and is found to be in the
range of 10-15 Earth masses ($M_{\bigoplus}$) (\citealp{Guillot2005}, see also \citealp{Movshovitz10} and \citealp{Piso15} for
updated discussion). The fundamental requirement of the core-accretion theory is to build up a planetary core that
reaches the critical mass within the PPD lifetime, which is typically a few million years \citep[e.g.][]{Haisch2001,Bell2013}.

Conventionally, it has been considered that building up a planetary core that reaches the critical mass is achieved by accreting planetesimals.
However, following a phase of runaway growth enabled by gravitational focusing \citep{Greenberg1978}, core
growth transitions to the phase of oligarchic growth and slows down substantially \citep{oligarchic}.
This is because a relatively massive core effectively stirs up the eccentricities of neighboring
planetesimals, greatly reducing the efficiency of gravitational focusing.
In the solar system, assuming a solid surface density comparable to the minimum mass solar nebula (MMSN,
\citealp{Weidenschilling1977b,MMSN}), forming the cores for Jupiter and Saturn by planetesimal accretion is only marginally
achievable. Towards larger distances beyond $5-10$ au, due to the increase in dynamical time and
the reduction in planetesimal surface density, the timescale to assemble a core of critical mass increases
rapidly with separation and well exceeds the disk lifetime \citep{Rafikov2004,Levison2010}. When applied to
the observed exoplanetary systems with giant planets at large separations, this standard model of core growth by planetesimal accretion simply fails.

Recently, it has been realized that, instead of accreting planetesimals, the growth of planetary cores
can grow much faster by accreting millimeter-centimeter sized \emph{pebbles} (\citealp{LJ12}, hereafter LJ12).
The presence of such pebbles is the outcome of grain growth, and they likely dominate the dust mass budget
as supported by millimeter dust continuum observations of PPDs \citep{Testi2003, Wilner2005, Rodmann06}.
Unlike planetesimals, pebbles of millimeter-centimeter size experience strong aerodynamic drag from the gas.
The importance of gas drag has already been realized in earlier works in the context of planetesimal
accretion \citep{Rafikov2004,KenyonBromley09}, where it helps accelerate core growth by damping planetesimal
eccentricities. Down to millimeter-centimeter sized pebbles, these particles are typically marginally coupled to the gas via gas drag in the outer regions of PPDs (in the sense that their stopping time $t_{s}$ is comparable to the dynamical time $\Omega^{-1}$). In this case, LJ12 showed that once the core mass exceeds some transition mass (defined in Equation (\ref{eq:Mtrans}) below), essentially all pebbles entering the core's Hill sphere, corresponding to the maximum gravitational reach of the core, can be accreted. This result agrees with an earlier numerical study by \citet{Johansen2010}, as well as with the analytical theory by \citet{Ormel2010} and test-particle integrations in hydrodynamic simulations by \citet{Morbidelli12}. The physics of pebble
accretion can also be considered as a special case of the wind-shearing effect in binary planetesimals
investigated by \citet{PeretsMurrayClay11}. With pebble accretion at maximum efficiency, the timescale
of core growth is substantially reduced, and the formation of giant planet cores at tens of au can be
accommodated well within the disk lifetime.

The original LJ12 model, as well as the analytical work of \citet{Ormel2010}, considers that pebble
accretion takes place in a laminar disk. However, the outer regions of PPDs are expected to
be turbulent due to the magnetorotational instability (MRI) \citep{ Balbus1991}. The overall level
of turbulence is still uncertain. On the one hand, we expect substantial damping of the MRI in the
midplane region of the outer disk due to strong non-ideal MHD effects, particularly ambipolar diffusion
(AD, \citealp{Bai2011}). On the other hand, the surface layer of the outer disk is likely to be fully
MRI turbulent, and its vertical extent depends on the depth that far-UV (FUV) photons can penetrate
\citep{PerezBecker2011}. Depending on the FUV penetration depth, which is uncertain and
may differ from system to system, the level of midplane turbulence can be enhanced by the MRI in the
surface FUV layer \citep{Simon_etal13b,Bai15}, a situation analogous to the conventional scenario of
layered accretion \citep{Gammie1996,Fleming2003}. Observationally, with uncertainties in disentangling
thermal and turbulent line broadening, evidence of disk turbulence has been inconclusive
\citep{Hughes2011,Guilloteau2012,Flaherty2015,Teague2016}. More recent results based on ALMA
observations may suggest that different disks
possess different levels of turbulence (e.g., \citealp{Flaherty2015,Teague2016}).

The presence of turbulence will have two effects on pebble accretion. First, it allows pebbles
to diffuse against vertical settling toward the midplane, and the level of turbulence determines
the thickness of the pebble layer $H_p$.
Second, the pebbles undergo turbulent stirring and their trajectories can substantially deviate
from analytical trajectories in a laminar disk.
Recent analytical models of planet formation that take into account the first effect
\citep{Morbidelli15,Ida16,Matsumura2017}, where 2D and 3D regimes are distinguished depending on whether
$H_p$ is smaller or larger than the pebble accretion radius $r_a$. In the former (2D) case,
all particles entering within $r_a$ of the core are assumed to be accreted. In the latter (3D)
case, only the fraction of pebbles located within the height of $r_a$ are counted.
However, it is unclear yet whether this treatment is sufficient, because analytical orbit
theory is violated due to turbulent stirring, and it is plausible that the efficiency of
pebble accretion can deviate from analytical predictions.

In this paper, we aim to test the efficiency of pebble accretion in the presence of the MRI
turbulence via numerical simulations. Our method is similar to LJ12, but our simulations
incorporate self-consistently generated MRI turbulence, and we control the strength of the
turbulence by incorporating AD as the main non-ideal MHD effect. We only
consider AD because we are mainly concerned with the formation of giant planets at large separations
at $\sim30$ au or beyond, where AD is the solely dominant non-ideal MHD effect \citep{Wardle2007,Bai11a}.
We emphasize that our work is not on the application of the pebble accretion theory to global
models of planet formation, as pursued by many authors (e.g., \citealp{Chambers14,Kretke2014,LJ14,Lambrechts2014,Bitsch15,Chambers16,Ida16,Sato16}).
Rather, we critically examine the theory of pebble accretion at microphysical level and test its
robustness under realistic MRI turbulence, so as to reassure researchers on the proper use
of analytical formulae for the rates of pebble accretion.

This paper is organized as follows. We describe the method and equations of our numerical
simulations in Section \ref{method}. In Section \ref{difsim}, we first present simulation results
on the vertical diffusion of particles, and then describe the setup for the simulations of pebble accretion.
We discuss our method of measuring and normalizing pebble accretion rates from our simulations in
Section \ref{results}. Our main results on the measured pebble accretion rate and particle kinematics
are presented in Section \ref{simresults}. In Section \ref{discussionandconclusion}, we summarize
the main results and discuss their applications.
 
\begin{table*}[]
  \begin{center}
    \caption{Characteristic Parameters of the Simulations in this Work.}
    \label{tab.simpar}
    \begin{threeparttable}
    \begin{tabular}{cccccccc}
      \toprule
      Simulation & Resolution & Box size($H$) & $M_c/M_T$ & $r_H (H)$ & $H_p(H)$ & $\beta_0$ & $Am$ \\
      \midrule
      hyd3 & $192/H$ & $1 \times 4 \times 1$ & $3 \times 10^{-3}$ & $0.1$ & $0.01$ & $\infty$ & - \\
      AD3 & $192/H$ & $2 \times 4 \times 1$ & $3 \times 10^{-3}$ & $0.1$ & S.C. & $12,800$ & $1$ \\
      idl3 & $192/H$ & $3 \times 6 \times 1$ & $3 \times 10^{-3}$ & $0.1$ & S.C. & $12,800$ & $\infty$ \\
      hyd2 & $96/H$ & $2 \times 4 \times 1$ & $3 \times 10^{-2}$ & $0.23$ & $0.01$ & $\infty$ & - \\
      AD2 & $96/H$ & $2 \times 4 \times 1$ & $3 \times 10^{-2}$ & $0.23$ & S.C. & $12,800$ & $1$\\
      idl2 & $96/H$ & $3 \times 4 \times 1$ & $3 \times 10^{-2}$ & $0.23$ & S.C.  & $12,800$ & $\infty$\\
      \bottomrule
    \end{tabular}
  \begin{tablenotes}
  \item \note All simulations have a particle number of 24,576 of each type per $H^3$, and assume $\Delta v_K = 	0.1 c_s$. ``S.C.'' stands for self-consistent.
  \end{tablenotes} 
  \end{threeparttable}
   \end{center}
\end{table*}

\section{Method} \label{method}
\subsection{Gas Dynamics}
We use the Athena magnetohydrodynamics (MHD) code \citep{Athena2008},
which is a higher-order Godunov code with constrained transport to enforce the
divergence-free constraint on the magnetic field, to perform local 3D MHD simulations under the shearing-sheet approximation \citep{GoldreichLyndenBell65}. We use a Cartesian coordinate system in a corotating frame located at a fiducial radius with Keplerian frequency $\Omega_K$. The dynamical equations are then written with $\hat{\mb{x}},\hat{\mb{y}},\hat{\mb{z}}$
denoting unit vectors pointing in the radial, azimuthal, and vertical
directions respectively, where ${\mb\Omega_K}$ is along the $\hat{\mb{z}}$ direction. With gas
density, gas velocity, and magnetic field denoted by $\rho_g, {\mb u}$ and
${\mb B}$, the MHD equations can be written as follows in this non-inertial
frame:
\begin{equation}\label{eq:gascont}
\frac{\pa\rho_g}{\pa t}+\nabla\cdot(\rho_g\mb{u})=0\ ,
\end{equation}
\begin{equation} \label{eq:gasmotion}
\begin{aligned}
&\frac{\pa\rho_g\mb{u}}{\pa t}+\nabla\cdot(\rho_g\mb{u}^T{\mb u}
+{\sf T}) \\ 
 =\rho_g\bigg[&2{\mb u}\times{\mb\Omega_K}+3\Omega_K^2x\hat{\mb{x}}+
2\Delta v_K\Omega_K\hat{\mb{x}}
- \nabla \Phi_P\bigg]\ ,
\end{aligned}
\end{equation}
\begin{equation}
\frac{\pa{\mb B}}{\pa t}=\nabla\times\bigg[{\mb u}\times{\mb B}
+\frac{({\mb J}\times{\mb B})\times{\mb B}}
{c\gamma\rho_{i}\rho_g}\bigg]\ ,
\label{eq:induction2}
\end{equation}
where ${\sf T}$ is the total stress tensor
\begin{equation}
{\sf T}=(P+B^2/8\pi)\ {\sf I}-\frac{{\mb B}^T{\mb B}}{4\pi}\ ,
\end{equation}
${\sf I}$ is the identity tensor, $P$ is the gas pressure, and $\Phi_P$ is the gravitational potential of the planetary core (see Section \ref{subsec.setup}). We assume an isothermal
equation of state $P=\rho_g c_s^2$, where $c_s$ is the isothermal sound speed.
The disk scale height is given by $H=c_s/\Omega_K$. In code units, we set
$c_s=\Omega_K=H=1$. Disk vertical gravity in the gas is ignored
and all of our simulations are vertically unstratified. This is because the length scale relevant to pebble accretion (the Hill radius of the planetary core) that we investigate in this paper is much smaller than $H$ (see Section \ref{subsec.setup}).
The initial gas density $\rho_0$ is hence uniform, and we set $\rho_0=1$ in code units.
Periodic boundary conditions are used in
the azimuthal and vertical directions, while the radial boundary conditions are
shearing periodic \citep{Hawley1995}.

The first and second terms on the right hand side of \eqref{eq:gasmotion} are standard shearing-sheet source terms of Coriolis force and tidal gravity, leading to velocity shear along $\hat{x}$. In practice, these terms are modified to subtract background shear motion using the orbital advection algorithm of \citet{StoneGardiner10} that improves the accuracy.

The third term corresponds to a constant force pointing radially outward (which we have newly implemented), representing a global radial pressure gradient in the disk. As a result,
the disk rotates more slowly than Keplerian by $\Delta v_K$. This sub-Keplerian rotation
is the source of the radial drift of dust particles, and sets the velocity scale of
dust particles approaching the planetary core (which rotates at Keplerian velocity and is stationary in our simulation frame). \footnote{This approach is equivalent to applying the opposite of the pressure gradient force on particles, as adopted in streaming instability simulations \citep{BaiStone10a}.  However, it has the advantage that, because we make the gas rotation sub-Keplerian, we can place a planet in Keplerian orbit at the center of the domain.} It is convenient to normalize $\Delta v_K$ by the sound speed $c_s$. We are mainly interested in the outer regions of PPDs, where pebble accretion is expected to play a dominant role in planetary core growth. In our simulations, we fix $\Delta v_K = 0.1c_s$ which is generally applicable in the outer PPDs.
\footnote{Assuming a standard MMSN disk, with $\Sigma(R) = 1700 (R/\textrm{AU})^{-3/2} $g cm$^{-2}$ and $T(R) = 280 (R/\textrm{AU})^{-1/2}$K, we have
\begin{equation}
\frac{\Delta v_K}{c_s} \approx 0.127 \left( \frac{R}{30AU} \right)^{1/4}.
\end{equation}
The observationally inferred disk surface distribution in the main body of the disk is shallower than the MMSN scaling ($\Sigma \sim R^{-1}$ instead of $R^{-3/2}$), which lowers $\Delta v_K / c_s$ to $\sim 0.1$ at 30 au. 
}.

The last term in \eqref{eq:induction2} represents AD, with
$\gamma$ denoting the coefficient of momentum exchange in ion-neutral
collisions, and $\rho_i$ being the ion density. In disks, AD can be most conveniently
parameterized by the Elsasser number, defined as
\begin{equation}
Am\equiv\frac{\gamma\rho_i}{\Omega_K}\ .\label{eq:Am}
\end{equation}
The value of $Am$ marks the
importance of AD: the ideal MHD regime corresponds to $Am\rightarrow\infty$, where
the magnetic field is frozen into the gas, while AD significantly affects gas dynamics when
$Am\lesssim10$ \citep{Bai2011}. In PPDs, AD is the dominant non-ideal MHD effect in the outer region ($R\gtrsim30$ au), where $Am\approx 1$ is found to be widely applicable \citep{Bai11a}.

We impose a net vertical magnetic field $B_0$ in our simulations. 
The strength of this net vertical field is measured by the plasma $\beta$:
\begin{equation}
\beta_0\equiv\frac{\rho_0c_s^2}{B_0^2/8\pi}、 ,
\end{equation}
which is the ratio of gas pressure to the magnetic pressure of the net vertical field.
Realistic simulations of PPD gas dynamics in the outer disk suggested that net vertical magnetic field is needed for the MRI turbulence to be sustained in the presence of strong AD \citep{Bai2011,Simon_etal13b}. It has also been found that
$\beta_0\sim10^4$ is needed to achieve the desired accretion rates consistent
with observations \citep{Simon_etal13b,Bai15}. In all MHD simulations described in this paper, we set $\beta_0=1.28\times10^4$.

We perform simulations with three different turbulence levels. First, we perform pure hydrodynamic simulations of pebble accretion for reference. These runs are labeled ``hyd" in Table \ref{tab.simpar}. We then perform non-ideal MHD simulations with $Am = 1$, which represent a realistic level of turbulence in the midplane regions for the outer PPDs, labeled by ``AD" in Table \ref{tab.simpar}. Finally, we perform simulations in ideal MHD, corresponding to an exaggerated level of turbulence, labeled by ``idl" in the Table. These simulations are described in detail in Section \ref{subsec.setup}.

\subsection{Particle dynamics}

Dust particles are included in our simulations using the particle module described in \citet{BaiStone10a}. The dust particles are treated as test particles that passively respond to the gas flow without exerting a backreaction on the gas, which allows us to cleanly separate the effect of the MRI turbulence from further complications due to particle feedback. We also note that \cite{LJ12} found that pebble accretion rate are essentially unaffected when the particles' backreaction is included. 
 
In the shearing-sheet framework, the
equation of motion for particle $i$ reads
\begin{equation}
\frac{d\mb{v}_i}{d
t}=2{\mb v}_i\times{\mb\Omega_K}+3\Omega_K^2x_i\hat{\mb{x}}
-\Omega_K^2z_i\hat{\mb{z}}-\frac{\mb{v}_i-\mb{u}}{t_{{\rm s}}}+{\mb a}_P, \ \label{eq:parmotion}
\end{equation}
where $v_i$ is particle velocity, $t_s$ is the particle stopping time characterizing the drag force, and ${\mb a}_P = - \nabla \Phi_P$ is the acceleration due to the planetary core (see Section \ref{subsec.setup}).
Note that unlike the gas, particles are not affected by the radial pressure gradient in the gas.  The resulting difference between gas and particle velocities causes the particles to feel gas drag, which leads to their radial drift. In addition, we have included vertical gravity for the particles, allowing them to settle toward the disk midplane (see Section \ref{subsec.verdiff}).

A particle's stopping time $t_s$ depends on properties of both the gas and the dust. In the low-density outer region of the disk, the relevant drag law is the Epstein law and $t_s = \rho_s a/\rho_g c_s$ \citep{Epstein}, where $\rho_s$ is the density of the dust material and $a$ is the particle radius, which is smaller than the mean free path of surrounding gas. 
It is most convenient to
use a dimensionless stopping time $\tau_s\equiv\Omega_K t_{s}$. Particles with
$\tau_s\ll1$ are strongly coupled to the gas, and particles with $\tau_s\gg1$ are loosely coupled
to it. Note that in the standard MMSN model, centimeter-sized particles have a dimensionless stopping time of $\tau_s \sim 0.1$ at 10 AU in the midplane, and $\tau_s \sim 1$ at  $\sim$40 au (see \citealt{Chiang2010} for a review).

In our pebble accretion simulation (see Section \ref{subsec.setup}), particles of seven different sizes are considered with stopping times ranging from $\tau_s = 10^{-2}$ to $\tau_s = 10$. In our simulations, we fix $\tau_s$ as to be constant for individual particles. This is because gas density is largely constant in the local disk midplane regions of interest, as in our unstratified simulations.

\section{Simulation setup} \label{difsim}
\subsection{Particle Vertical Diffusion Simulation} \label{subsec.verdiff}

Because of vertical gravity, dust particles settle toward the disk midplane, balanced by turbulent diffusion. Therefore, before proceeding to pebble accretion simulations, we first conduct MRI simulations without a planetary core to determine the vertical profile of dust particles under different levels of turbulence in steady state.
This will be used to initialize pebble accretion simulations to be described in the next subsection.

Two simulations are conducted, in either ideal MHD or non-ideal MHD with $Am=1$.
Both runs use the same box size of $H\times4H\times H$ in $x, y$ and $z$
dimensions, resolved by $192\times384\times192$ cells. The use of such a high
resolution first guarantees that the most unstable MRI wavelength $\lambda_m$ is
resolved. In ideal MHD, we have $\lambda_m\approx9.18\beta_0^{-1/2}H\approx0.081H$
for $\beta_0=12800$. This corresponds to $\sim16$ cells per scale height, which is reasonably well resolved.
For non-ideal MHD with $Am=1$, $\lambda_m$ is roughly doubled \citep{Wardle99,Bai2011},
and we expect the MRI to be very well resolved in this run. In addition, the high
resolution we use here is necessary to properly resolve the Hill sphere of the planetary
core in our pebble accretion simulations. 

We first run simulations without particles to time $t=120\Omega_K^{-1}$ when the MRI is
expected to fully saturate. We then inject nine
types of particles, each characterized by a constant dimensionless stopping time
$\tau_s$ spanning from $0.01$ to $100$ with two particle types per decade in
$\tau_s$. We inject 28,800 particles per type, whose spatial distribution follows a
Gaussian profile $\propto\exp{(-z^2/2H_{p0}^2)}$, where $H_{p0}$ is an initial guess of
particle scale height ranging from 0.003 for the most loosely coupled particles to 0.3 for the most strongly coupled. We further wait for another $\gtrsim20$ orbits for particles
to interact with turbulence before starting to measure particle vertical profiles.

The profiles can be well fitted by a Gaussian, characterized by particle scale height $H_p$,
which is expected from the balance between vertical settling and turbulent diffusion. 
The value of $H_p$ is related to the strength of turbulence, and particle stopping time $\tau$.
Theoretically, we expect \citep{YoudinLithwick07}
\begin{equation} \label{eq.YL}
H_p\approx\sqrt{\frac{D_{g,z}}{\Omega\tau_s}}=\sqrt{\frac{\alpha_z}{\tau_s}}H,
\end{equation}
where $D_{g,z}\equiv\alpha_zc_sH$ is the turbulent diffusion coefficient in the gas.

\begin{figure}[htb]
  \centering	
    \includegraphics[width = 0.5\textwidth]{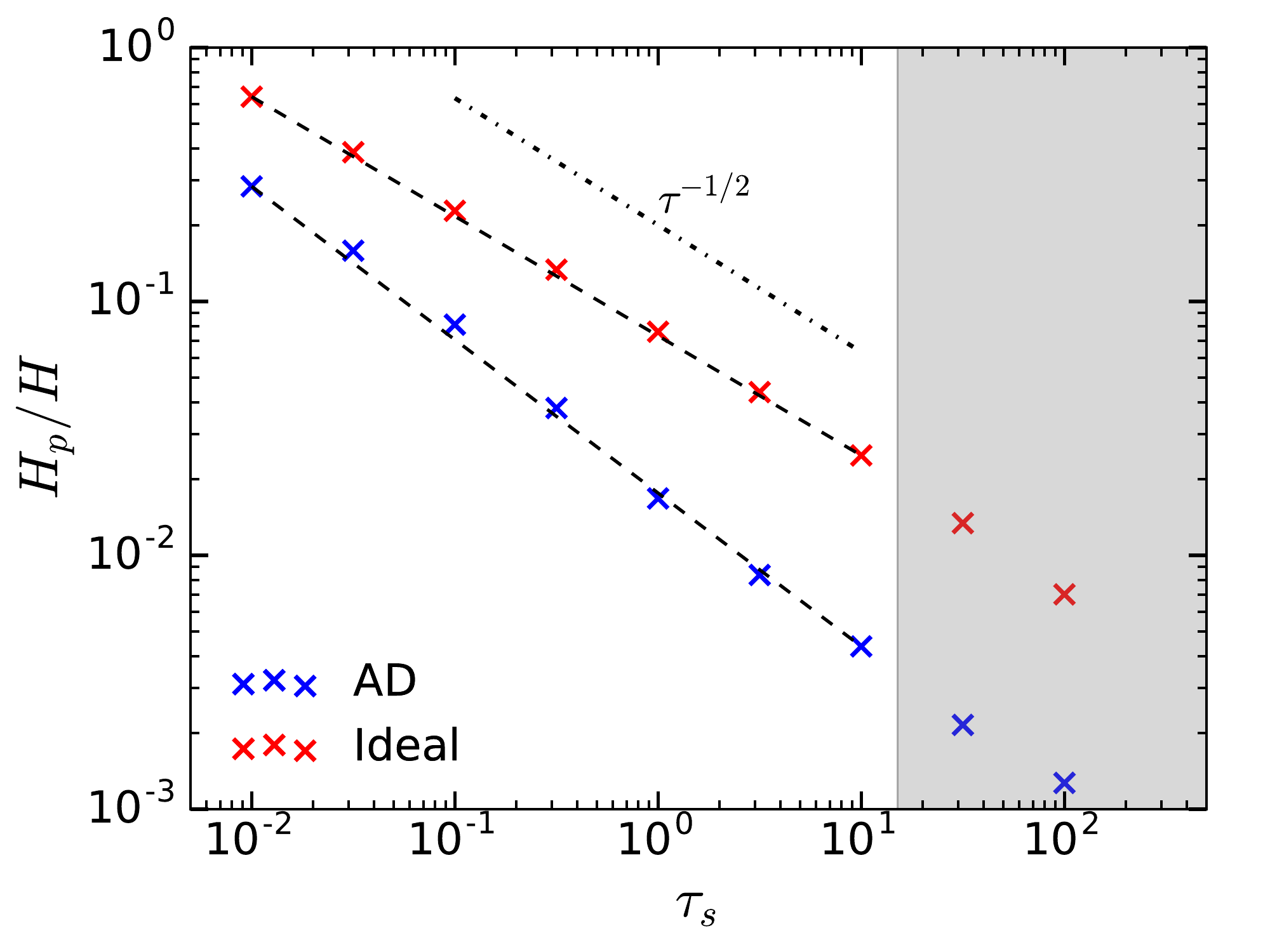}  
  \caption{Particle scale heights $H_p$ as a function of dimensionless particle stopping time $\tau_s$. The blue and red cross symbols represent results from our AD and ideal MHD simulations, respectively. The black dashed lines simply connect the scale heights of $\tau_s = 0.01$ and $\tau_s = 10$ particles, which fit the intervening simulation data well. The black dotted line indicates a power-law slope of $-1/2$. The data points for $\tau_s >10$ particles (in the shadowed region) are not used in our accretion simulations, but are included for test purposes.}
  \label{Fig.Hp}
\end{figure}

The results for the normalized particle scale heights $H_p/H$ in our simulations as a function of $\tau_s$ are shown in Figure \ref{Fig.Hp}.
The dependence of $H_p$ on $\tau_s$ in both simulations nicely follows a power law with index close to (but not exactly) $-1/2$, in reasonable agreement with theoretical expectations, as well as with previous simulation results \citep{Carballido_etal11,Zhu_etal15}. The particle scale heights obtained here (based on the dashed lines shown in Figure \ref{Fig.Hp}) will be used for initializing the pebble accretion simulations. By fitting the results using (\ref{eq.YL}), we find vertical diffusion coefficient $\alpha_z = 7.8\times 10^{-4}$ for the AD simulation and $\alpha_z = 4.4\times 10^{-3}$ for the ideal MHD simulation.

For reference, we also measure the $\alpha$ parameter for disk angular momentum transport \citep{Shakura1973}, which is obtained by evaluating the time- and volume-averaged sum of the Maxwell stress and Reynolds stress, normalized by thermal pressure: 
\begin{equation}
\alpha = \langle \frac{-B_x B_y+\rho v_x v_y}{\rho c_s^2} \rangle.
\end{equation}
We find $\alpha = 6.8\times 10^{-4}$ in the AD simulation and $\alpha = 0.045$ in the ideal MHD simulation. We note that the difference in $\alpha_z$ in between the ideal MHD and AD runs is much smaller than their difference in $\alpha$. This is related to the longer turbulent correlation time in the AD case, as studied in \citet{Zhu_etal15}. More specifically, we find the mean square of the vertical turbulent velocity to be $\langle v_z^2/c_s^2\rangle =3.0\times10^{-4}$ and $\langle v_z^2/c_s^2\rangle = 0.0121$ in the AD and ideal simulations, respectively. This is consistent with the results of \citet{Zhu_etal15} where
$D_{g,z}\sim 3-4\Omega^{-1}\langle v_z^2 \rangle$
in the AD case and
$D_{g,z}\lesssim 0.5\Omega^{-1}\langle v_z^2 \rangle$
in the ideal MHD case.

\subsection{Pebble Accretion Simulations: Setup and Parameters} \label{subsec.setup}

We conduct pebble accretion simulations with three different levels of turbulence as mentioned before, and two different planetary core masses $M_c$, with a total of six runs. Their simulation parameters are listed in Table \ref{tab.simpar} and described in more detail below.

We choose the planetary core mass to be such that it is sufficiently massive for pebble accretion to proceed in the most efficient ``Hill regime" \citep{LJ12} while not too massive to open a gap in the disk.
The core mass is best normalized by the thermal mass \citep{LinPapaloizou93}, defined as
\begin{equation}
M_T=\frac{c_s^3}{G\Omega_K}\approx 160M_\oplus\bigg(\frac{R}{30\rm AU}\bigg)^{3/4}\ ,
\end{equation}
where the number in the approximate equality is obtained based on the temperature profile of the MMSN disk model around a protostar of solar mass. With this definition, the Hill radius of the planetary core is given by
\begin{equation}
r_H=\bigg(\frac{M_c}{3M_T}\bigg)^{1/3}\ .
\end{equation}
\citet{LJ12} defined the ``transition mass" for pebble accretion as
\begin{equation}\label{eq:Mtrans}
M_t=\frac{\Delta v_K^3}{G\Omega_K}\approx
0.160M_\oplus\bigg(\frac{R}{30\rm AU}\bigg)^{3/4}\bigg(\frac{\Delta v_K}{0.1c_s}\bigg)^3\ ,
\end{equation}
where they showed that beyond this core mass, particles with stopping time $\tau_s\sim0.1-1$ passing the core within the Hill radius can be accreted by the core (which defines the ``Hill regime").
For our choice of $\Delta v_K=0.1c_s$, we have $M_t=10^{-3}M_T$. In our simulations, we choose the core masses to be $M_c=3\times10^{-3}M_T$ and $3\times10^{-2}M_T$, and
we attach the numbers $3$ and $2$ to the names of corresponding runs. These core masses
guarantee that pebble accretion occurs in the Hill regime. We avoid choosing an even larger core mass, which would notably affect the bulk flow structure and potentially open gaps in the vicinity of the core \citep{Dong_etal11b,Ormel13}, leading to additional effects that make it more difficult to isolate the role of turbulence.

The Hill radii for the two choices of core mass are $r_H=0.1H$ and $r_H=0.23H$, respectively. In the former case, we choose the grid resolution to be 192 cells per $H$, so that $r_H$ is resolved by about $20$ cells. In the latter case, we reduce the resolution to $96$ cells per $H$ and the Hill radius is resolved by about an equal number of cells.

In the simulations, the planetary core is placed at the center of the simulation box, with its gravitational potential given by
\bgeq
\Phi_p=-GM_c\frac{r^2+3R_s^2/2}{(r^2+R_s^2)^{3/2}}\ ,
\edeq
where $r$ is the distance to the core, and $R_s$ is the softening length. The smoothing function is accurate to fourth-order at $r\gg R_s$ while it avoids divergence at $r<R_s$ (e.g., \citealp{Dong_etal11b}). The softening lengths $R_s$ for the gas are chosen to be about $0.02H$ for most of the simulations, except for runs idl3 and AD3, which use $0.01H$.
We use half the gas softening length for particles,\footnote{In Athena, external gravity on the gas is evaluated from the finite difference of the gravitational potential over neighboring cells. Therefore, $R_s$ needs to span across a few cells to guarantee smoothness of the gas flow. On  the other hand, we use direct gravitational force to integrate the particles, which allows particles to feel deeper gravitational potential without being affected by grid resolution.}, which is only slightly larger than the grid scale and is much smaller than $r_H$ (by a factor $\sim10$). This ensures that the process of pebble accretion is largely unaffected by the smoothed gravitational potential.
For runs idl3 and AD3, using a greater softening length would make some particles in our smallest size bin ($\tau_s=0.01$) be stripped from the artificially shallow potential well after they are accreted.\footnote{We originally chose $R_s=0.02H$ for run AD3, and found that $\tau_s=0.01$ particles are mostly stripped from the core even enough they are counted as being accreted (see Section 4.2). We hence reduce $R_s$ to $0.01H$ which resolves this issue. Nevertheless, Figures other than Figure 4 (focusing on larger particles) are made using the data from run AD3 with $R_s=0.02H$, where the change in $R_s$ makes little difference.}

We also note that the Bondi radius of the core $R_B=GM_c/c_s^2=(M_c/M_T)H\ll H$ is roughly at grid resolution for $M_c=3\times10^{-3}M_T$, and it is resolved by about three cells for $M_c=3\times10^{-2}M_T$. While this is not sufficient to properly resolve
the gas flow in the vicinity of the planetary core as well as its atmosphere (e.g., \cite{Ormel15}), our choice of particle smoothing length $R_s$
guarantees particles of the sizes considered here ($\tau_s\gtrsim0.01$) are strongly bound to the core once accreted,.

At the beginning of our pebble accretion simulations,
we first gradually introduce the gravity from the planetary core over a period of $\sim10$ orbits ($60\Omega_K^{-1}$) without adding particles. This allows the gas to adapt to the presence of the core. For all MHD runs, we further run the simulations to time $t=180\Omega_K^{-1}$ to allow the MRI turbulence to fully develop. We then start to
inject particles. We use seven particle types spanning dimensionless stopping time from
$\tau_s=0.01$ to $\tau_s=10$, and each particle type contains $24,576$ particles per $H^3$. For MHD simulations, particles are injected uniformly in the horizontal domain, and
their vertical distribution follows the Gaussian profile $\propto\exp{(-z^2/2H_p^2)}$
with particle scale heights $H_p$ for each particle type derived from previous
diffusion simulations. For hydrodynamic simulations, all the particles are initialized to be in the midplane. All particles are initialized with velocities following the solution in the laminar disk given by Equations \eqref{eq.vr} and \eqref{eq.vphi} below. As in \citet{LJ12}, we use open/outflow boundary conditions for particles (not gas) in the radial and azimuthal directions so that any particles that leave the box will no longer re-enter from the other side.
Correspondingly, all particles transiently pass the core in the simulations, and either get captured or get lost.

In the MHD runs, because of the stochastic nature of the MRI turbulence, we repeat the particle injection process multiple times while we continue running the MRI simulations, which allows us to improve the statistics by sampling different realizations of the MRI turbulence.
Note that the timescale for particle passage through the core is of the order of
$H/\Delta v_K\sim10\Omega_K^{-1}$. In practice, we repeat particle injection for six cycles for the most MRI simulations, with each cycle spanning $\Delta t_{\rm cycle}=30\Omega_K^{-1}$, except for run idl3, where we cover 10 cycles with each cycle spanning $\Delta t_{\rm cycle}=24\Omega_K^{-1}$.

In our simulations, we adopt simulation box sizes from $H\times4H\times H$ to $3H\times6H\times H$ (see Table \ref{tab.simpar}). They are chosen in the main to enable a larger fraction of particles in the simulation box to be accreted, and in particular, we use the relatively extended domain in the radial dimension in order to access particles carried by large turbulent eddies on scales of $\sim H$.

\begin{figure*}
  \centering
 \subfigure[MHD simulation with ambipolar diffusion, particle size $\tau_s=0.1$.]
 {\label{fig.AD3tau0.1_mov}
    \includegraphics[width = 0.48\textwidth]{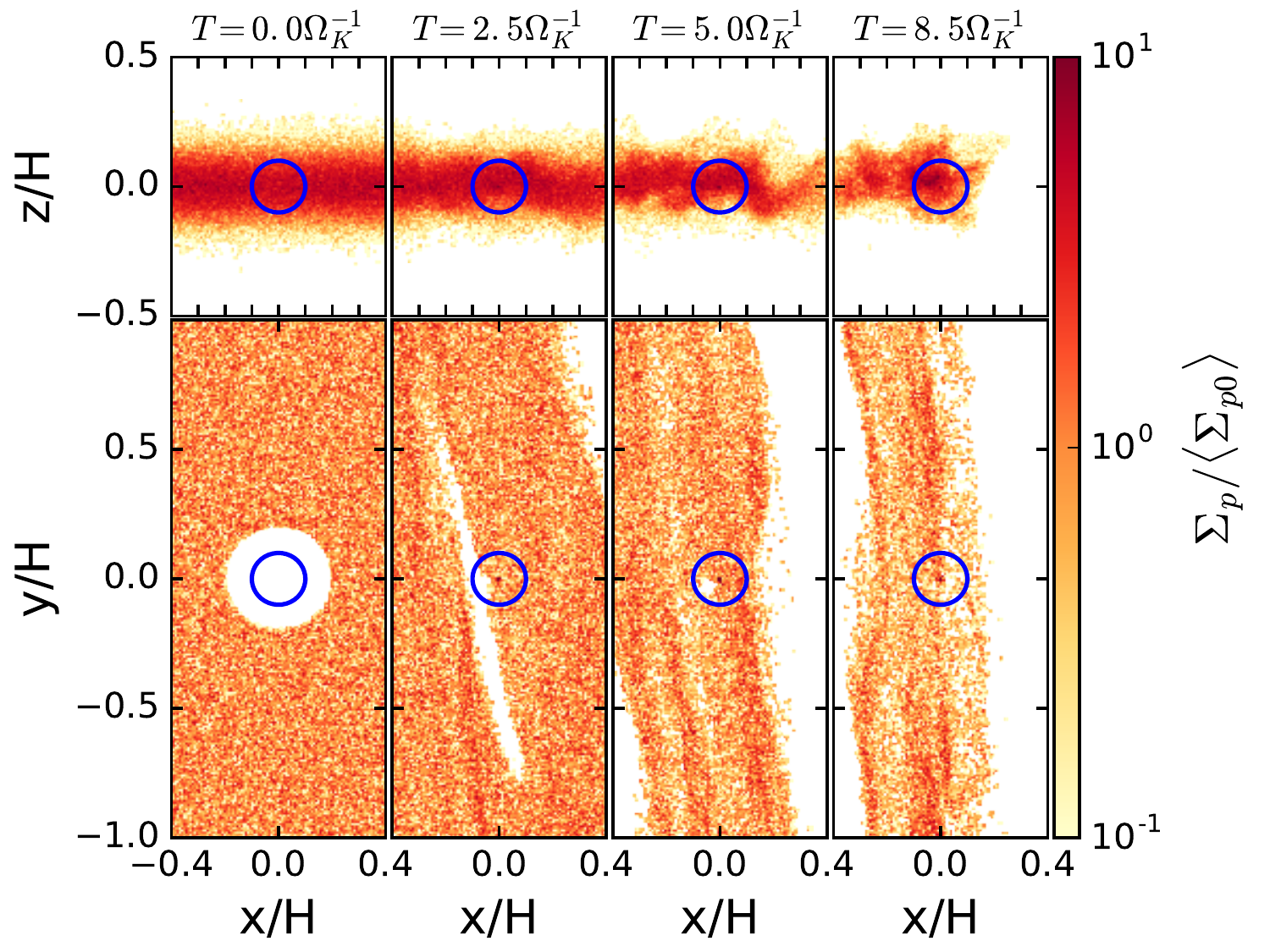}  
 }
 \subfigure[MHD simulation with ambipolar diffusion, particle size $\tau_s=1$.]
 {\label{fig.AD3tau1_mov}
    \includegraphics[width = 0.48\textwidth]{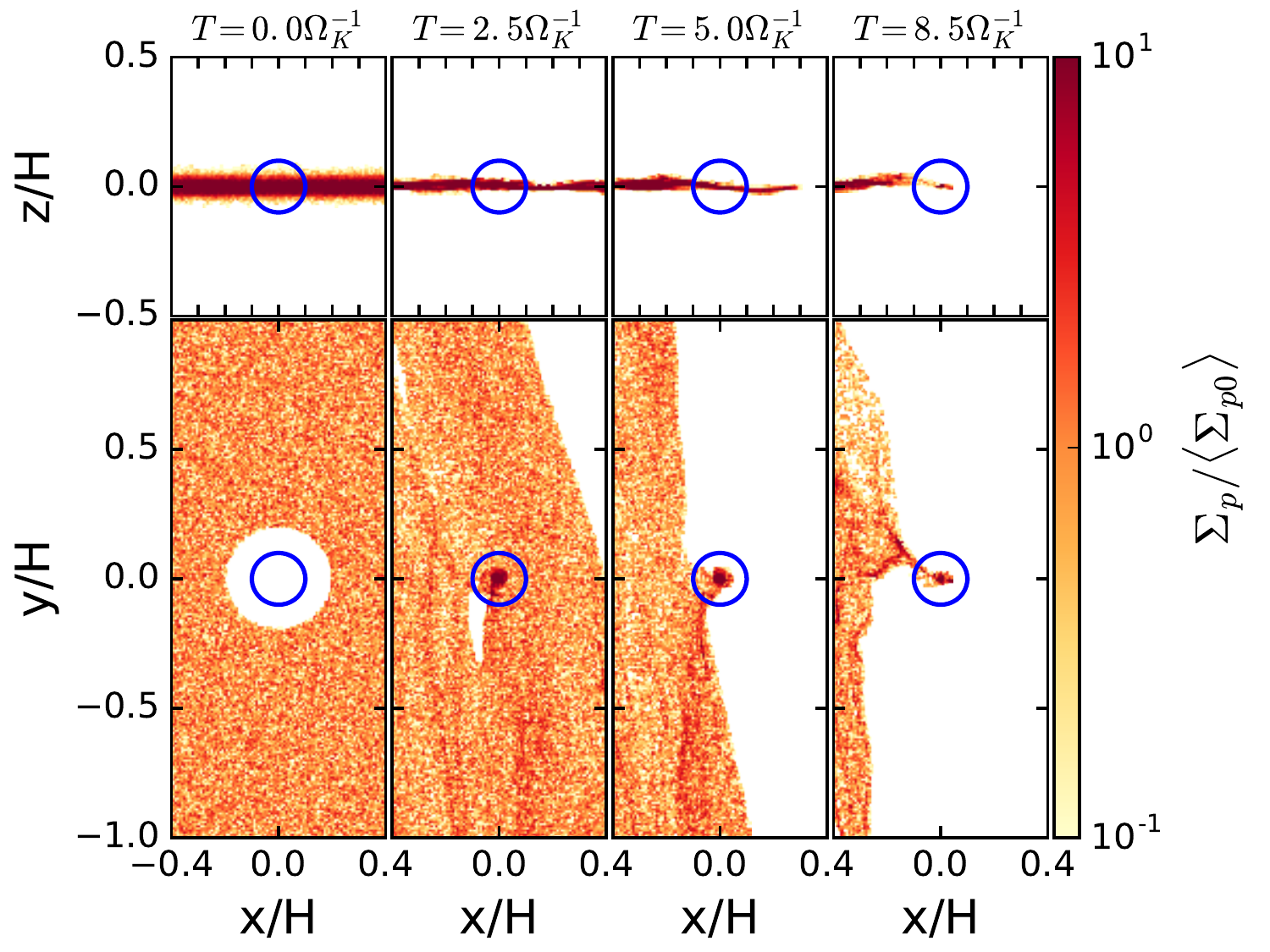}  
 }
 \subfigure[Ideal MHD simulation, particle size $\tau_s=0.1$.]
 {\label{fig.fig.idl3tau0.1_mov}
    \includegraphics[width = 0.48\textwidth]{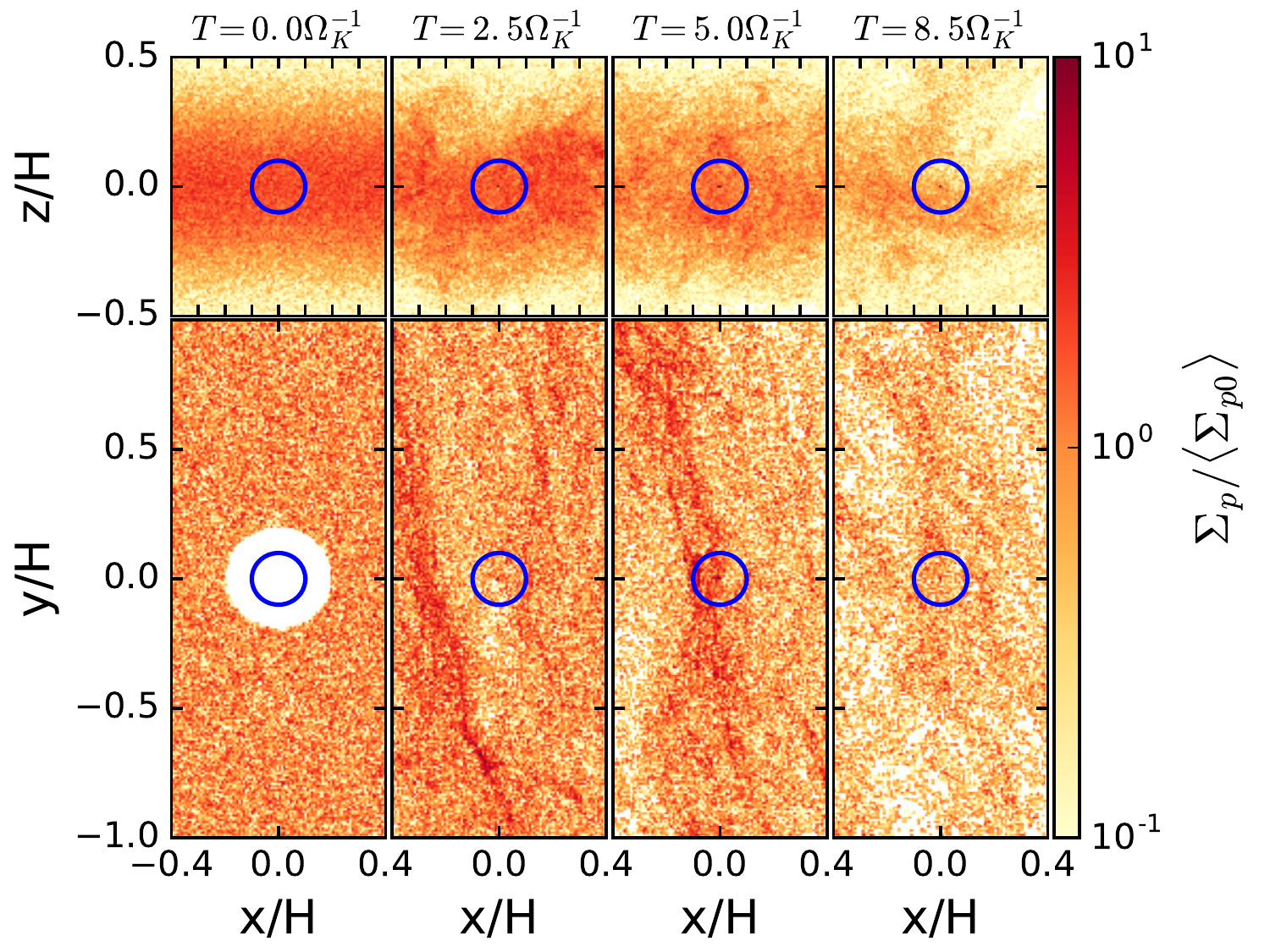}  
 }
 \subfigure[Ideal MHD simulation, particle size $\tau_s=1$.]
 {\label{fig.fig.idl3tau1_mov}
    \includegraphics[width = 0.48\textwidth]{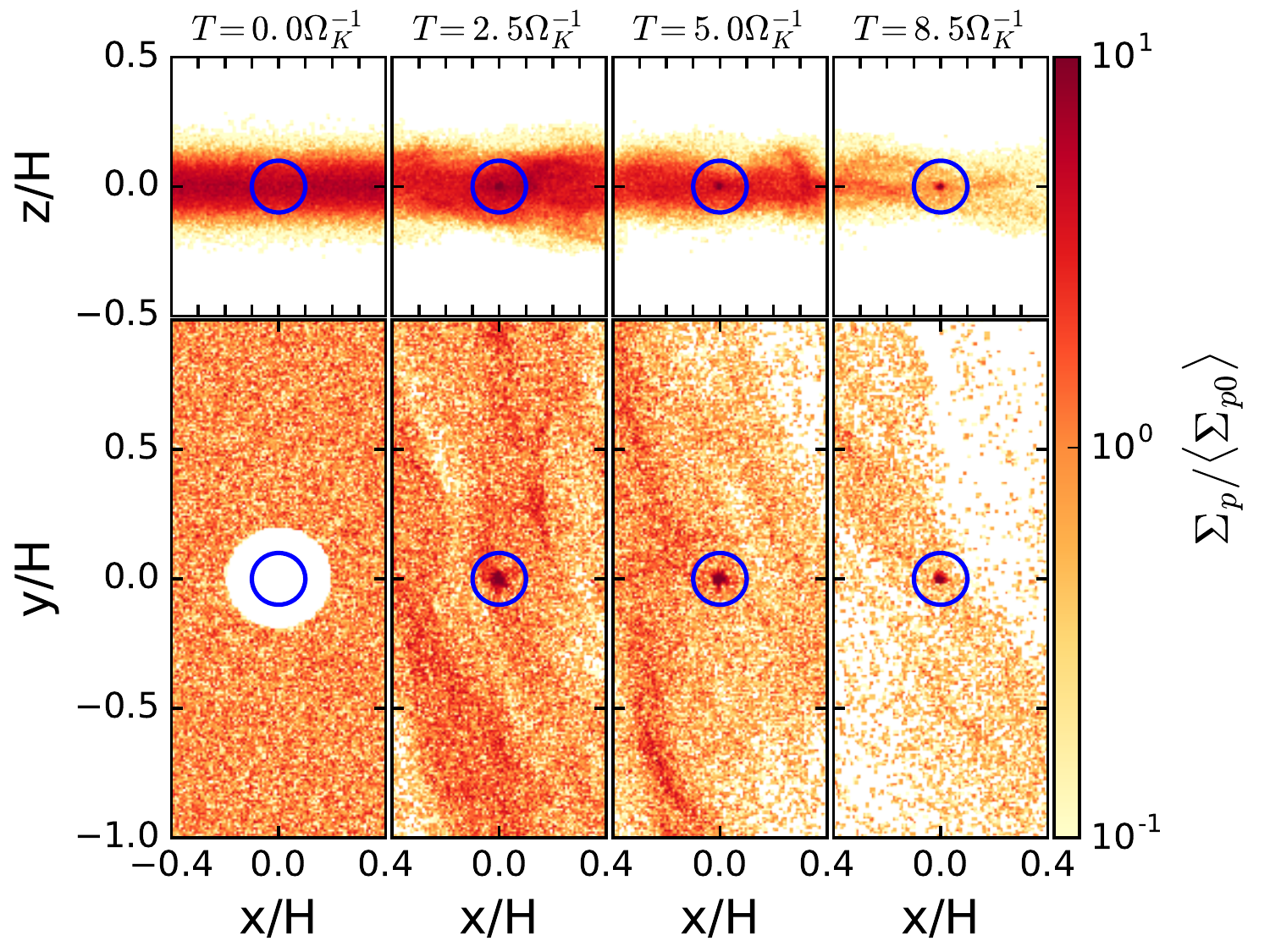}  
 }

 \caption{\label{fig.mov}Snapshots of projected particle column densities in the x-y and x-z planes for $\tau_s=0.1$ and $\tau_s = 1$ particles in the central part of our ideal MHD and AD simulations, for core mass $\mu_c = 3\times 10^{-3}$. The snapshots are taken from the first cycle of particle injection in each simulation run. The blue circles in the center of the boxes indicate the boundary of the Hill sphere. Particles initially located within $2r_H$ from the core are removed in the lower panels to better demonstrate the accretion process of particles outside the Hill sphere.}
\end{figure*}

\section{Measuring the Rate of Pebble Accretion} \label{results}

Our main goal is to measure the rate of pebble accretion from our simulations. In this section, we first briefly describe the general appearance of our simulations, and then discuss measurements of accretion rate.

\subsection{Overview of Simulations}

Figure \ref{fig.mov} shows snapshots of projected particle densities from initial to later ($T = 8.5\Omega_K^{-1}$) times in the central part of ideal MHD and AD simulations for particle sizes $\tau_s = 0.1$ and $1$. The snapshots correspond to the first cycle of particle injection.
In the absence of turbulence, particles drift relative to Keplerian orbits in both
radial and azimuthal directions because of sub-Keplerian motion of the gaseous
disk. For particles with dimensionless stopping time $\tau_s$, their drift velocity relative to the local Keplerian velocity is given by
\begin{equation} \label{eq.vr}
v_r = -2\frac{\tau_s}{\tau_s^2+1}\Delta v_K,
\end{equation}
\begin{equation} \label{eq.vphi}
v_{\phi} = -\frac{1}{\tau_s^2+1}\Delta v_K,
\end{equation}
\citep{Weidenschilling1977,Nakagawa1986}. In our simulation frame, their velocity also includes Keplerian shear:  
\begin{equation} \label{eq.vsh}
v_{sh}(x) = -\frac{3}{2}\Omega_K x\hat{y}\ .
\end{equation}
In addition, particles are subject to the gravitational attraction of the core, as well as random kicks from the MRI turbulence.

In simulations with AD, turbulence is relatively weak. Unless particles are strongly coupled, they largely follow trajectories in the laminar flow to zeroth order. This is best seen in the snapshots for $\tau_s=1$ particles shown in Figure \ref{fig.AD3tau1_mov}, where bulk particle motion is dominated by radial drift and shear. Most particles are captured by the core as they pass by, leading to density enhancement at the center, while others exit the box. The overall process is relatively brief, and takes much less than $10\Omega_K^{-1}$ because we run out of particles in the simulation box. For $\tau_s=0.1$ particles, while the overall process is similar, evolution (and hence accretion rate) is slower because the radial drift speed is lower for more strongly coupled particles. We also see that the distribution of particles are more blurry, again because these particles are more strongly coupled to the gas and hence are more strongly affected by turbulence.

Snapshots from ideal MHD simulations show dramatic differences. For both $\tau_s=0.1$ and $\tau_s=1$ particles shown in Figures \ref{fig.fig.idl3tau0.1_mov} and \ref{fig.fig.idl3tau1_mov}, the motion of all particles is strongly perturbed by turbulence, with their velocity very effectively randomized. A large fraction of particles linger around the core for much longer time. While accretion still occurs, it is much less obvious what fraction of particles is accreted.

\begin{figure*}
  \centering
 \subfigure[Particle size $\tau_s=0.1$.]
 {\label{fig.kabs_t_shift_tau0.1}
    \includegraphics[width = 1.0\textwidth]{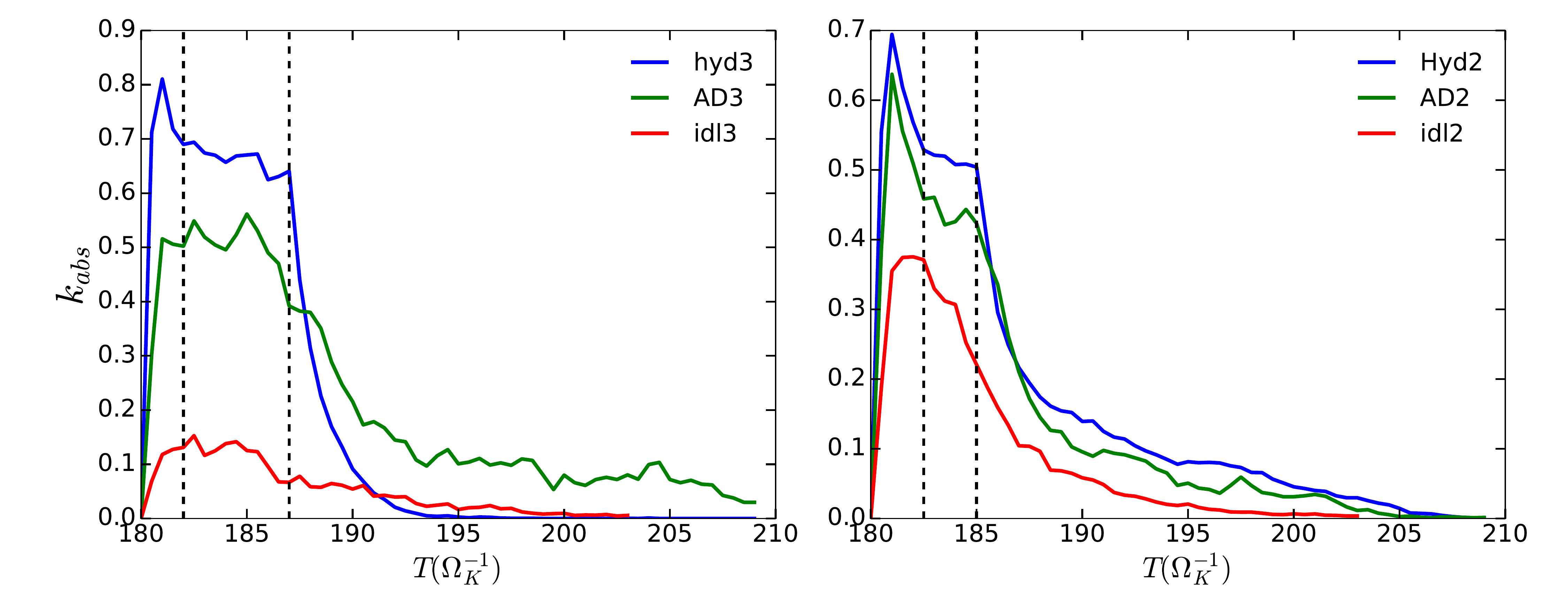}  
 }
 \subfigure[Particle size $\tau_s=1$.]
 {\label{fig.kabs_t_shift_tau1}
    \includegraphics[width = 1.0\textwidth]{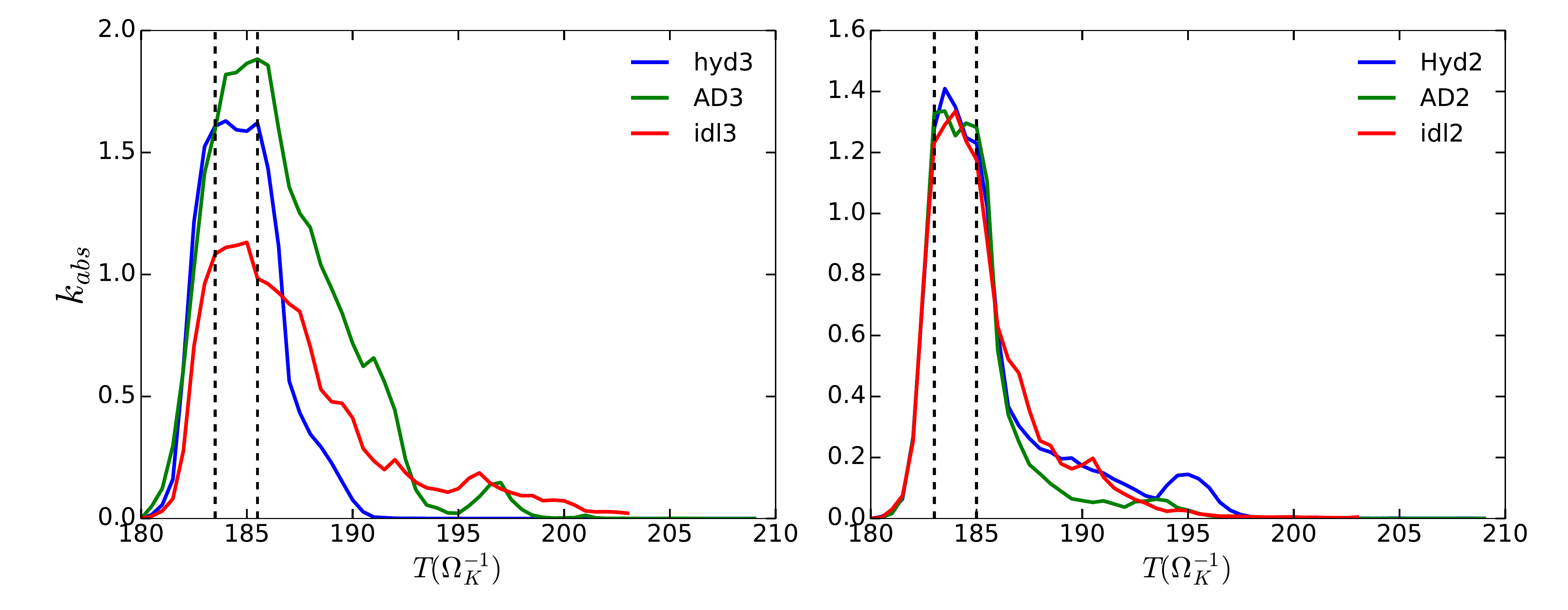}  
 }

 \caption{ \label{fig.kabs_t_shift} Instantaneous accretion rate coefficient $k_{abs}$ for $\tau_s=0.1$ (top) and $\tau_s=1$ (bottom) particles in all our simulations as a function of time (averaged over all cycles of particle injection). The time interval between the black dashed lines in each panel marks the steady accretion stage we identified from the hydrodynamic simulations. In the measurement, we have removed particles whose initial total energy is below the threshold energy so as to reduce the influence from initial conditions. Particles are injected into the simulation at $T=180\Omega_{K}^{-1}$, after the turbulence has had time to fully develop.}
\end{figure*}

Additionally, we see that in the AD simulation with $\tau_s=1$ (Figure \ref{fig.AD3tau1_mov}), the particle scale height is well below $r_H$, and hence accretion pebble accretion proceeds essentially in a 2D manner: the entire particle layer is potentially within the reach of the core, and most of the physics can be understood by restricting particle orbits to the disk midplane. In other cases, however, with particle scale height $H_p\gtrsim r_H$, pebble accretion is inherently 3D. One might think that only particles located within $z=\pm r_H$ can potentially be accreted. However, because particle scale height is maintained by turbulent diffusion, individual particles follow complex trajectories and can travel through very different vertical heights during their passage of the core. As a result, it is not obvious what fraction of particles can be captured, and whether one can use theories developed in the 2D case \citep{Ormel2010,LJ12} to predict the rate of pebble accretion. We will address this question through more detailed analysis.

\subsection{Measurement of Accretion Rates} \label{accrate_measurement}

To measure the instantaneous particle accretion rate, we first discuss the criterion to for a particle to be counted as being accreted. We adopt an energy criterion based on the total energy of individual particles relative to the planetary core, defined as
\begin{equation}
E_b = E_p + E_k = -GM_c\frac{r^2+3R_s^2/2}{(r^2+R_s^2)^{3/2}} + \frac{1}{2}(v_x^{'2}+v_y^{'2}+v_z^{'2})\ ,
\end{equation}
where $v'_x, v'_y, v'_z$ are particle velocities with Keplerian shear motion subtracted. For accreted particles, the total energy is negative and gradually decreases as the particle spirals into the core due to the energy dissipation. For the escaped particles, the total energy will remain positive until the end of the simulation. There is a lower limit to the total energy, which is set by the softening length, given by setting $r=0$ and $E_k=0$:
\begin{equation}
E_{p0} =  -\mu_c \left. \frac{r^2+3R_s^2/2}{(r^2+R_s^2)^{3/2}}\right|_{r=0} = -\frac{3\mu_c}{2R_s}\ ,
\end{equation}
corresponding to particles falling into the center of the gravitational potential well. Because we do not resolve the Bondi radius of the planet, the depth of the potential well $E_{p0}$ is limited by the softening length $R_s$, associated with grid resolution. 

We count a particle as being accreted when its total energy falls below an energy threshold, which is taken to be $3/4$ of $E_{p0}$:
\begin{equation}\label{eq:threshold}
E_{b0} = \frac{3}{4}E_{p0} = -\frac{9GM_c}{8R_s}\ .
\end{equation}
This corresponds to the gravitational potential energy at $r\approx 2R_s/3$, at which the softening of the gravitational potential becomes significant.
With our choice of the softening length, we confirm that in almost all runs, the vast majority of particles of all sizes that meet this energy threshold end up being accreted into the core in all our simulations.
The only exceptions involve $\tau_s=0.01$ particles in the runs idl3 and AD3, where some ($\lesssim20\%$) fraction of these particles that meet our criterion (\ref{eq:threshold}) are later kicked out of the core by the MRI turbulence. 
This is largely owing to the softening of the gravitational potential used in our simulations.
In reality, we still expect many of these particles to fall into the core, although the accretion rates measured based on criterion (\ref{eq:threshold}) likely represent an overestimate. 

We then determine the steady accretion stage to calculate the accretion rate. In Figure \ref{fig.kabs_t_shift}, we show the measured time sequence of instantaneous particle accretion rates for $\tau_s=0.1$ and $\tau_s=1$ particles (averaged over all cycles). Because particles are initially released without random velocities, it takes a short period of time ($\sim\tau_s^{-1}$) before they catch up with turbulent motion in the gas, as well as for them to respond to the gravity of the core. In addition, we expect a steady state to be achieved after particles initially located right outside of the Hill sphere reach the core, which roughly takes a time of around $r_H/\Delta v_K\sim\Omega^{-1}$. Overall, it should take up to a few $\Omega_K^{-1}$ (longer for more loosely coupled particles) for the particle accretion rate to reach a steady state. To better illustrate how a steady state is achieved, we have removed particles whose initial total energies are below the threshold energy \eqref{eq:threshold}. From Figure \ref{fig.kabs_t_shift} we see that approximate steady-state accretion is indeed achieved for essentially all runs; it is exhibited as a plateau following a rapid rise in accretion rate. The steady-state period is relatively brief (a few $\Omega_K^{-1}$); it is followed by a rapid drop in particle accretion rate as we run out of particles available within the simulation box.

We use the accretion rate curves of hydrodynamic simulations as the standard of reference, and mark the beginning and ending times of these plateaus for each core mass and particle size as the time interval for steady accretion. The same interval is used for the corresponding AD and ideal MHD simulations, and we can see from Figures \ref{fig.kabs_t_shift_tau0.1} and \ref{fig.kabs_t_shift_tau1} that, in general, the time intervals chosen from pure hydrodynamic simulations match reasonably well for AD and ideal MHD simulations. We measure the steady-state particle accretion rate by averaging the accretion rates over the marked time intervals.

Although we include particles with stopping time as large as $\tau_s=10$ in our simulations, it takes much longer for these particles to respond to turbulence, during which time a large fraction of them have left our simulation box. Therefore, we consider our measured accretion rates for these particles to be less reliable. Moreover, since grains typically continue to grow up to about centimeter size \citep{Birnstiel2011}, where they have $\tau_s<10$ in the bulk of an MMSN disk, we expect our study to cover most of the relevant parameter space for PPDs. 

\subsection{Normalization of Measured Accretion Rates}

\subsubsection{Absolute Accretion Rate Coefficient} \label{kabs.meth}
  
A natural scale to normalize particle accretion rate in the Hill regime is
\citep{LJ12}
\begin{equation}
\dot{M}_{2D} = 2\Sigma_p r_H v_H
=3\Sigma_p \Omega_K r_H^2, \label{Mdot2D}
\end{equation}
where $\Sigma_p$ is the particle surface density,
and $v_H \equiv(3/2)\Omega_Kr_H$ is the Hill velocity, characterizing the velocity with which particles approach the core at the Hill radius. Effectively, it is the characteristic accretion rate in the 2D case, with accretion radius $\sim r_H$.
We define the \emph{absolute accretion rate coefficient} $k_{\rm abs}$ as
\begin{equation}
k_{\rm abs} = \frac{\dot{M}_{\rm sim}}{\dot{M}_{2D}} = \frac{\dot{M}_{\rm sim}}{3 \Sigma_p \Omega_Kr_H^2},
\end{equation}
where $\dot{M}_{sim}$ is the accretion rate calculated in our simulations. 
This coefficient is independent of particle size, and provides a direct, dimensionless measure of absolute particle accretion rates as a function of $\tau_s$.

\subsubsection{2D/3D-modified Accretion Rate Coefficient} \label{kmod.meth}

When the thickness of the particle layer reaches or exceeds $r_H$, as mentioned earlier, the accretion process is inherently 3D. The effect is noted and distinguished in \citet{Morbidelli15}. Here, we provide another way to normalize the measured particle accretion rate that takes into account the finite thickness of the particle layer, which transitions smoothly from the 2D regime to the 3D one as $H_p$ varies.

Our procedure is a direct generalization of the 2D normalization factor, by dividing the particle layer into 2D sheets at individual heights, applying \eqref{Mdot2D}, and then integrating over height:
\begin{equation}
\begin{split}
\dot{M}_{\rm mod} &= \int_{-r_H}^{r_H}2\rho_p(z)R_H(z) v_{\rm sh}(z)dz\\
&= \int_{-r_H}^{r_H}3\rho_p(z)\Omega_K(r_H^2 - z^2)dz\ .
\end{split}
\end{equation}
where $R_H(z)\equiv\sqrt{r_H^2 - z^2}$, $v_{sh}(z)\equiv(3/2)\Omega_KR_H(z)$, $\rho_p(z)=\rho_{p0}e^{-z^2/(2H_p^2)}$ is the local particle density, and $\rho_{p0}=\Sigma_p/(\sqrt{2\pi} H_p)$ is the midplane particle density. Note that \cite{Morbidelli15} used the mass flux $\dot{M}_F$ of pebbles due to radial drift as the normalization factor, which is dependent on particle size. Our choice of normalization factor is independent of $\tau_s$, which is more convenient for comparing the pebble accretion rate among different particle sizes.

Similar to Section \ref{kabs.meth}, we now define the \emph{2D/3D-modified accretion rate coefficient}, $k_{\rm mod}$, as
\begin{equation}\label{eq:kmod}
k_{\rm mod} = \frac{\dot{M}_{\rm sim}}{\dot{M}_{\rm mod}}\ .
\end{equation}
This coefficient better characterizes the {\it intrinsic efficiency} of pebble accretion when the particle layer is puffed up, and allows us to directly compare simulations with different levels of turbulence.

 \begin{figure*}
  \centering
 \subfigure[Core mass $\mu_c = 3\times 10^{-3}$.]
 {\label{fig.kabs_3_errbar_batch}
    \includegraphics[width = 0.48\textwidth]{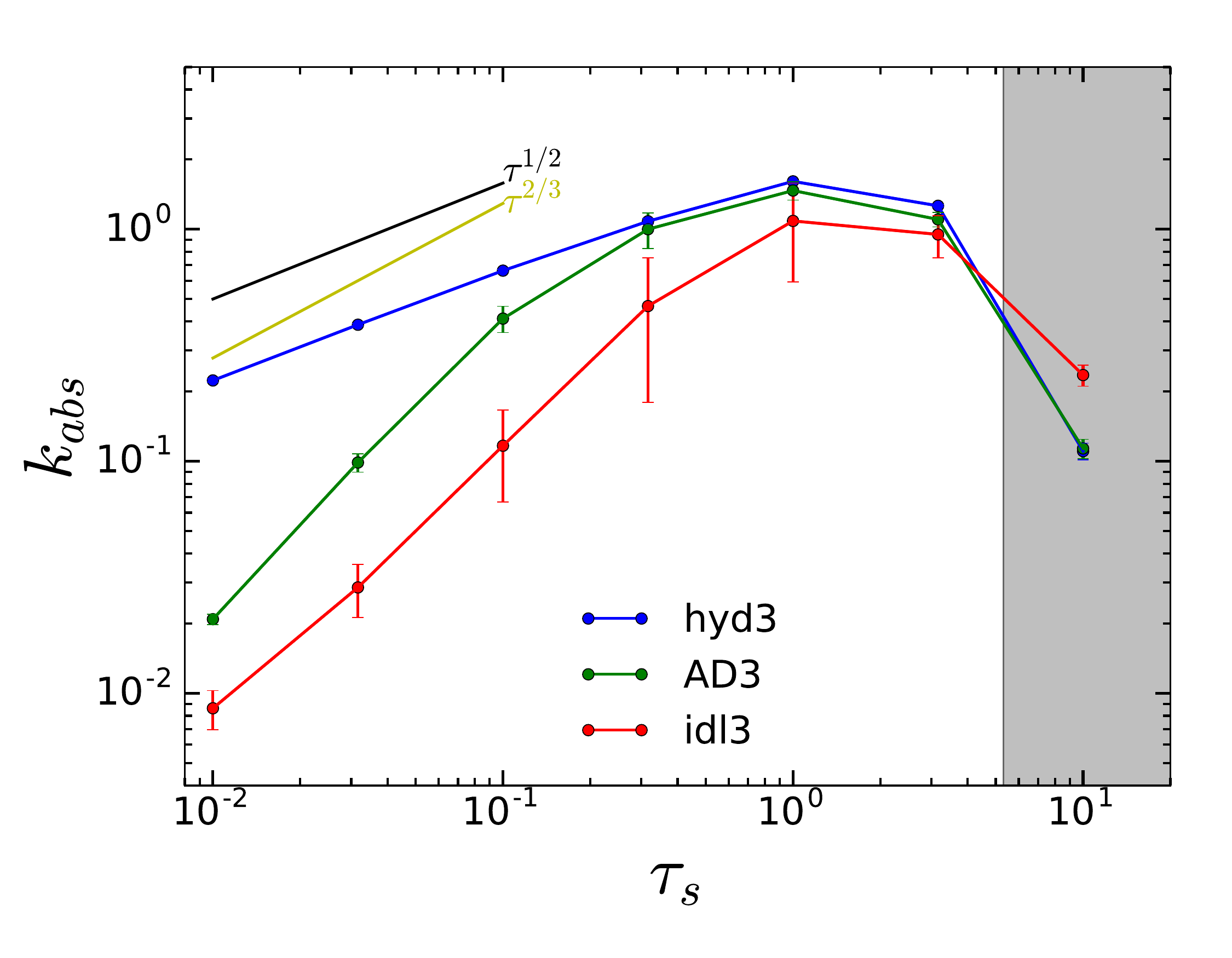}  
 }
 \subfigure[Core mass $\mu_c = 3\times 10^{-3}$.]
 {\label{fig.kmod_3_errbar_batch}
    \includegraphics[width = 0.48\textwidth]{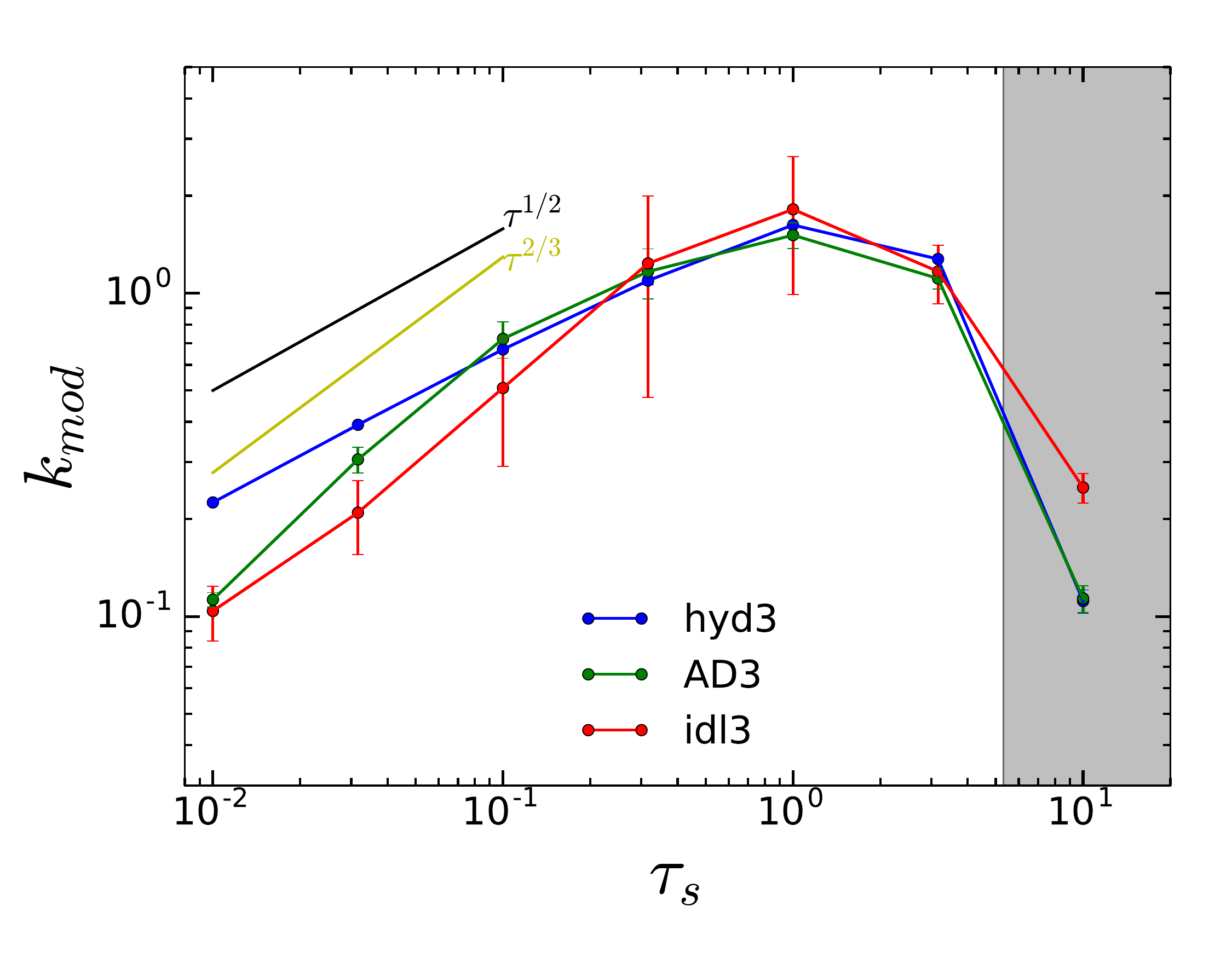}  
 }
 \subfigure[Core mass $\mu_c = 3\times 10^{-2}$.]
 {\label{fig.kabs_2_errbar_batch}
    \includegraphics[width = 0.48\textwidth]{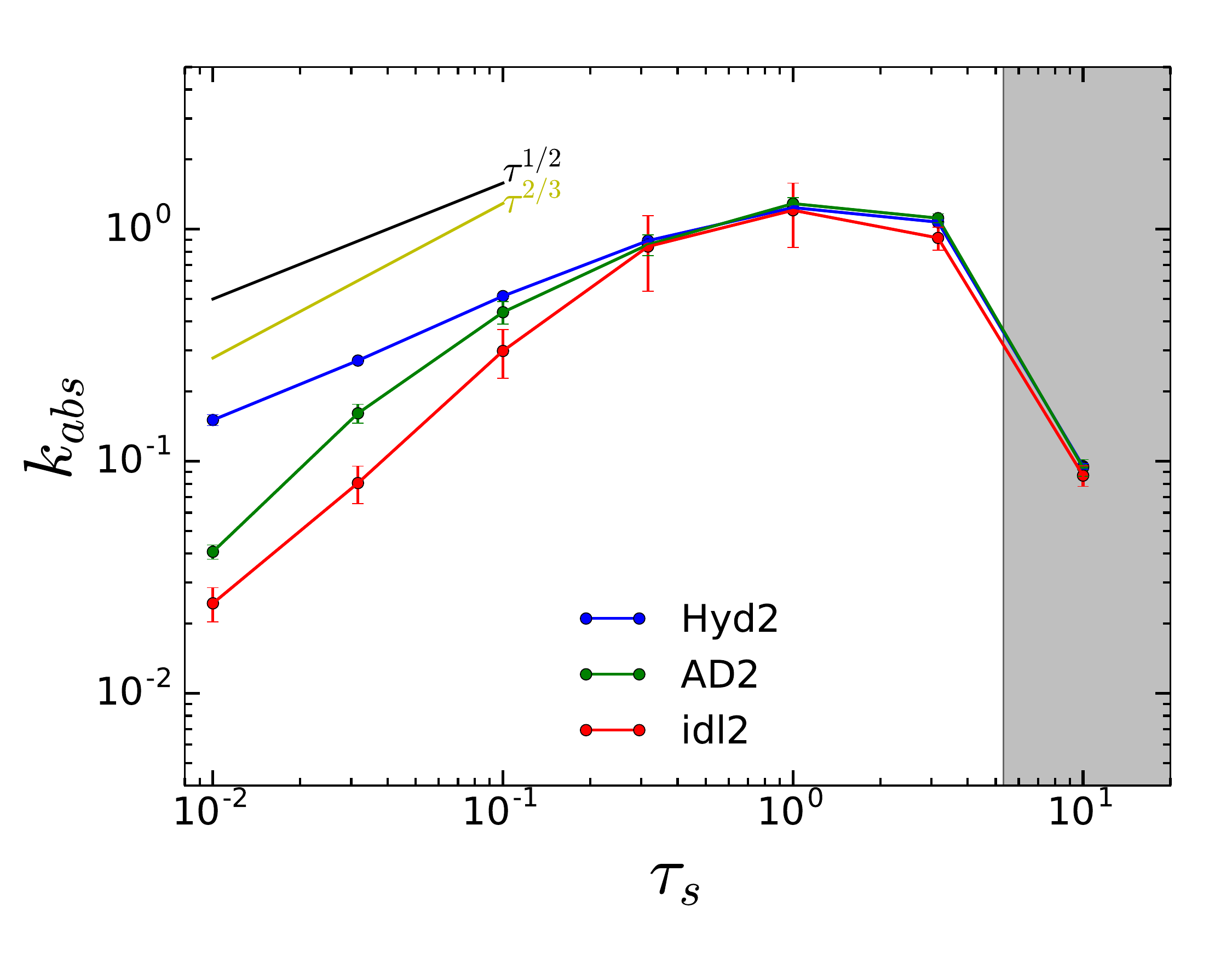}  
 }
 \subfigure[Core mass $\mu_c = 3\times 10^{-2}$.]
 {\label{fig.kmod_2_errbar_batch}
    \includegraphics[width = 0.48\textwidth]{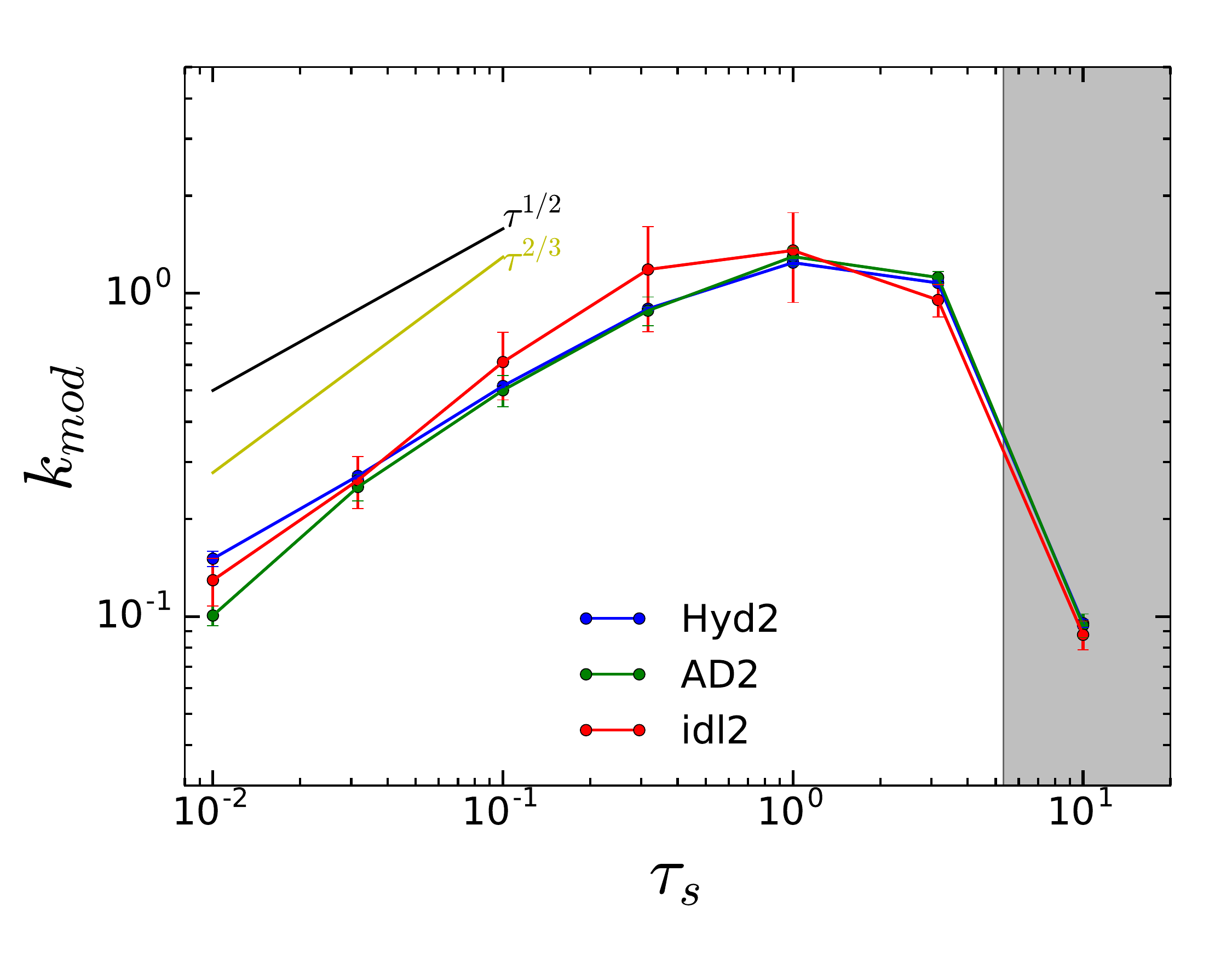}  
 }

 \caption{\label{fig.k_coefficient_errbar_batch}
 Absolute (left) and modified (right) accretion rate coefficients measured from all our simulations as a function of dimensionless stopping time $\tau_s$. The black and yellow solid lines represent the power-law scalings $\sim \tau_s^{1/2}$ and $\sim \tau_s^{2/3}$, corresponding to the theoretical expectations of particle accretion rates in the drift and Hill regimes, respectively. Results for large particles (in the gray region) are less reliable because of their longer response time to turbulence. The error bars are estimated from the statistical uncertainties of different particle injection cycles.}
\end{figure*}

\subsection{Theoretical Expectations}\label{ssec:theory}

Theoretically, the particle accretion rate is largely determined by the accretion radius $r_a$, the maximum impact radius below which particles can be accreted by the core. \cite{LJ12} qualitatively derived scaling relations between $r_a$ and particle stopping time $\tau_s$. In the so-called ``drift regime" with core mass below transition mass $M_t$ \eqref{eq:Mtrans}, the accretion radius is given by (see also \citealp{Baruteau16})
\begin{equation}\label{eq-drift}
r_a\sim\tau_s^{1/2}\bigg(\frac{M_c}{M_t}\bigg)^{1/6}r_H\propto\tau_s^{1/2}\ ,
\end{equation}
for $\tau_s\lesssim1$. In the ``Hill regime" that we consider here, these authors find
\begin{equation}\label{eq-hill}
r_a\sim\tau_s^{1/3}r_H\ ,
\end{equation}
again for $\tau_s\lesssim1$. We provide scalings derived from these approximate asymptotic relations for comparison with our results. 
\footnote{Our drift and Hill regimes altogether correspond to the "settling regime" of \citet{Ormel2010}, where accretion of small particles is primarily governed by gas drag. \citet{Guillot2014} and \citet{Homann2016} further considered the "hydrodynamic regime". While turbulence is found to enhance the accretion rates of particles in this regime to approach the geometric limit, it applies only to very small particles whose stopping time is below the timescale for the flow to pass by the embryos, which is very different from what we focus in our paper.}

The rate of pebble accretion also depends on the velocity $v_{\rm rel}$ at which particles approach the core. For $\tau_s \lesssim 1$, the particles are well coupled to the gas, so that $v_{\rm rel}$ is roughly the relative velocity between the core and the gas: 
\begin{equation}
v_{\rm rel}={\rm max}[\Delta v_K, v_{\rm sh}(r_a)]\ .
\end{equation}
In the drift and Hill regimes, $v_{\rm rel}$ is dominated by drift ($\Delta v_K$) or shear ($v_{\rm sh}$), respectively. Since $r_a\lesssim r_H$ and $v_{\rm sh}\propto r_a$, even with $M_c>M_t$, drift can dominate for sufficiently strongly coupled particles.

The corresponding accretion rate in the 2D case is approximately given by
\begin{equation}
\dot{M}\approx 2\Sigma_p r_av_{\rm rel}\ .
\end{equation}

\begin{figure*}
  \centering
 \subfigure[Core mass $\mu_c = 3\times 10^{-3}$, particle size $\tau_s=0.1$.]
 {\label{fig.traj_all_m3tau0.1}
    \includegraphics[height = 0.28\textheight]{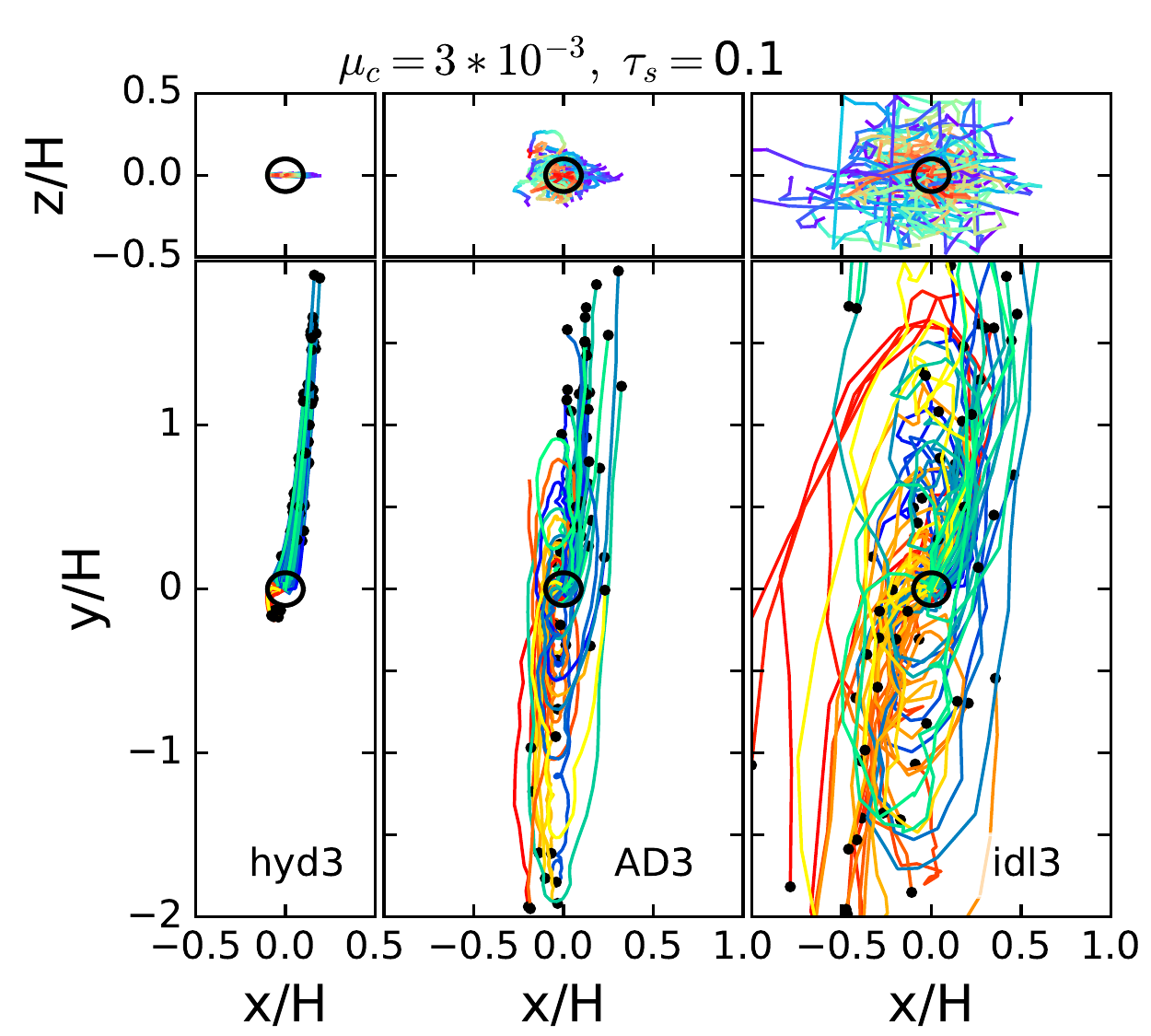}  
 }
 \subfigure[Core mass $\mu_c = 3\times 10^{-2}$, particle size $\tau_s=0.1$.]
 {\label{fig.traj_all_m2tau0.1}
    \includegraphics[height = 0.28\textheight]{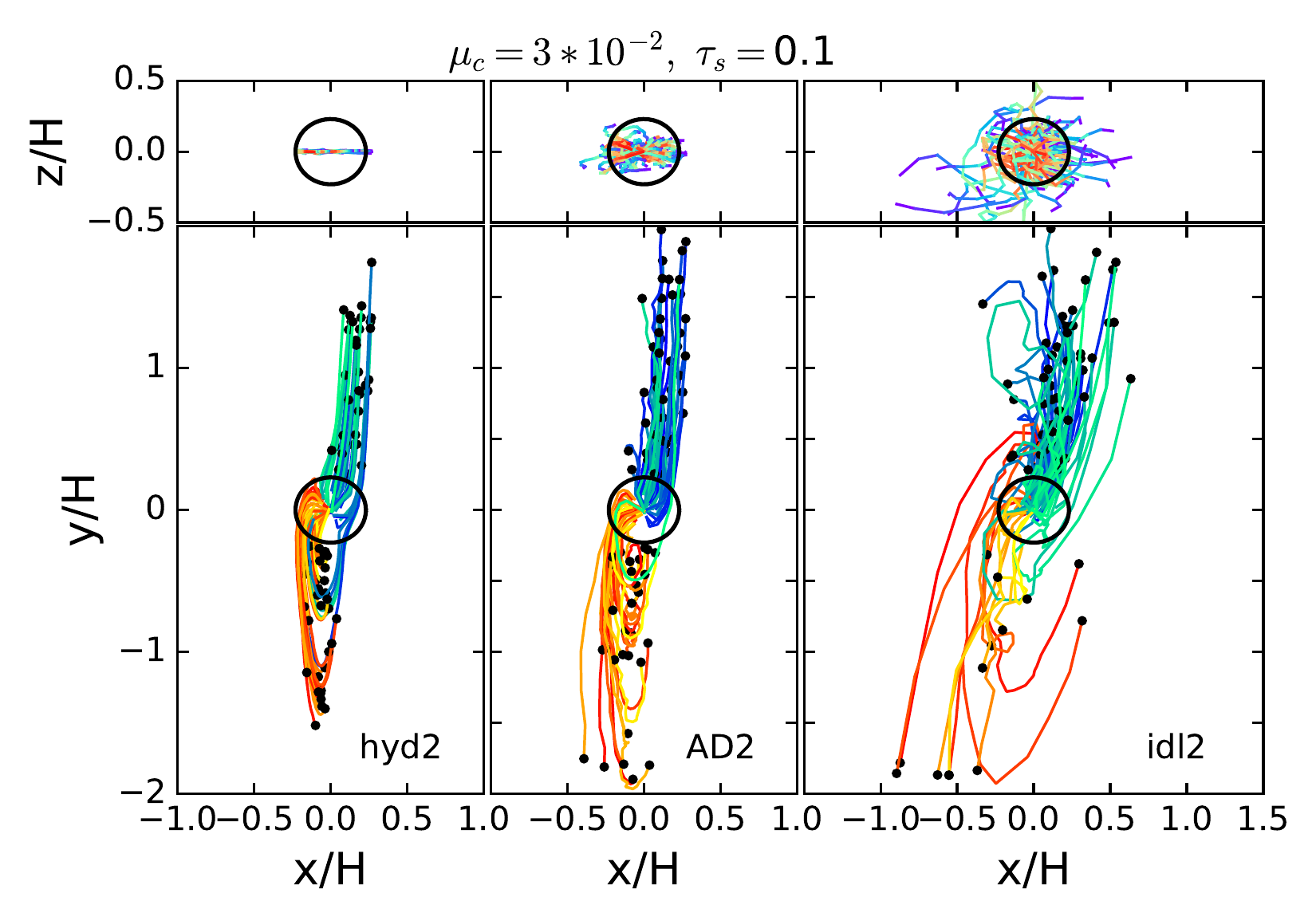}  
 }
  \subfigure[Core mass $\mu_c = 3\times 10^{-3}$, particle size $\tau_s=1$.]
 {\label{fig.traj_all_m3tau1}
    \includegraphics[height = 0.28\textheight]{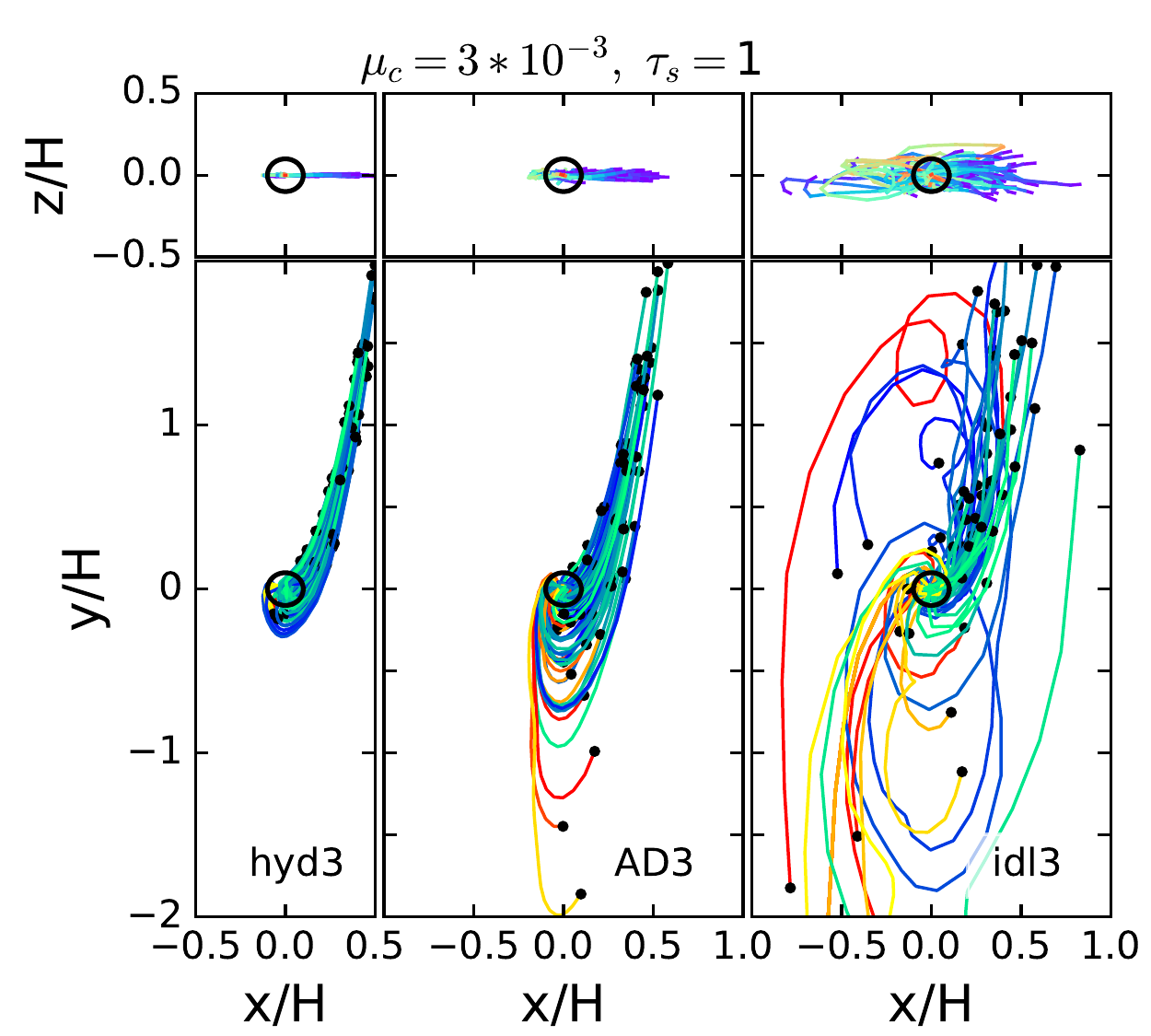}  
 }
  \subfigure[Core mass $\mu_c = 3\times 10^{-2}$, particle size $\tau_s=1$.]
 {\label{fig.traj_all_m2tau1}
    \includegraphics[height = 0.28\textheight]{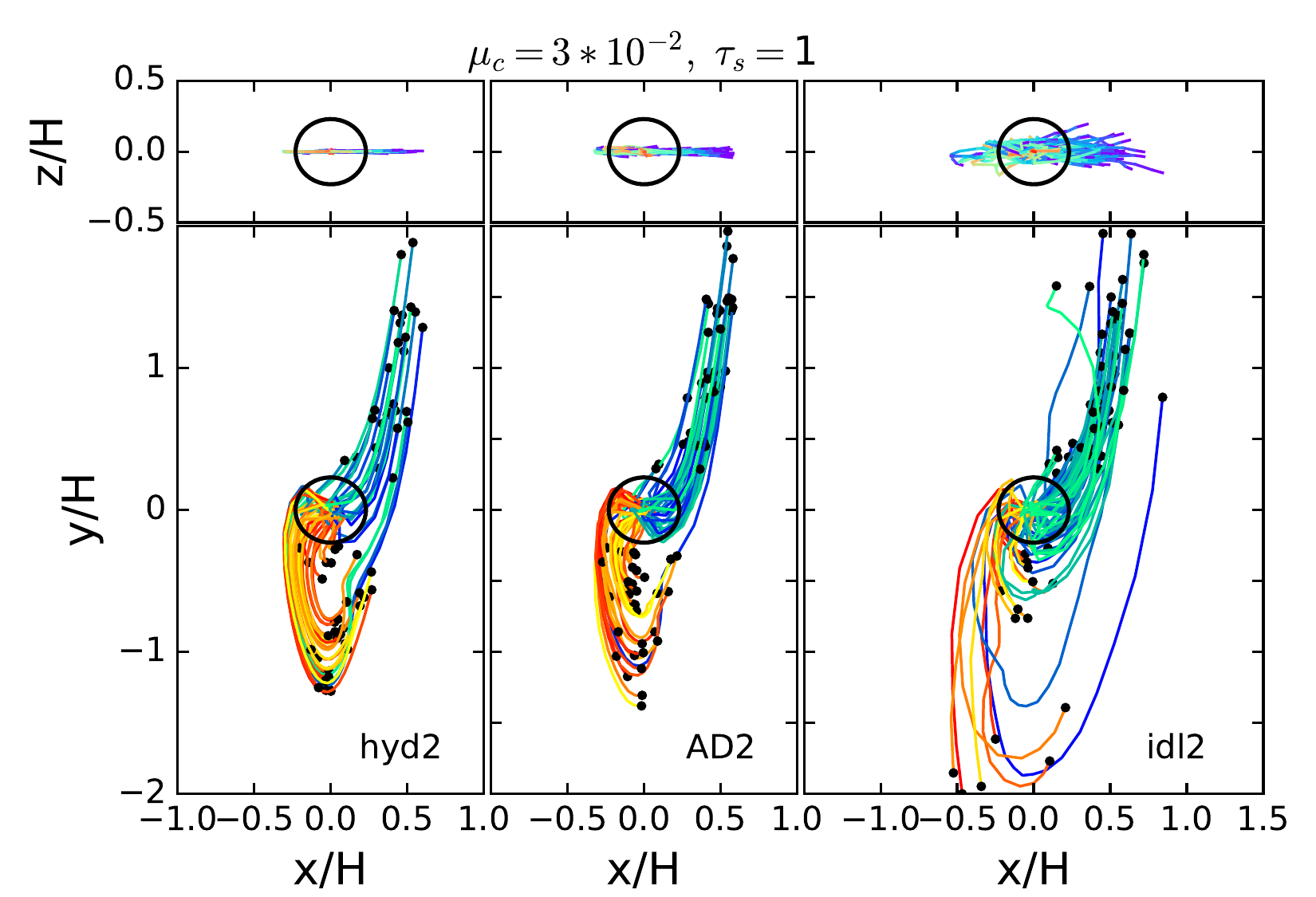}  
 }

 \caption{ \label{Fig.Traj} Representative trajectories of particles that are captured by the core in simulations with particle sizes of $\tau_s = 0.1$ and $\tau_s = 1$. The black circles at the center of each panel show the boundary of the Hill sphere, and the black dots in the x-y plane (lower panels in each subfigure) indicate the initial positions of each particle. Trajectories in the x-y plane are color-coded based on their initial positions. Trajectories in the x-z plane (upper panels in each subfigure) are colored from cold (blue) to warm (red) according to time.}
\end{figure*}
Equations (\ref{eq-drift}) and (\ref{eq-hill}) with appropriate choices for $v_{rel}$ imply that $\dot{M}\propto\tau_s^{1/2}$ in the drift regime, and $\dot{M}\propto\tau_s^{2/3}$ in the Hill regime. Note that the 2D accretion rate directly applies for a laminar disk. Generalization to 3D can be made using the 2D/3D-modified normalization factor described in Section \ref{kmod.meth}. If the {\it efficiency} of pebble accretion is similar in 2D and in 3D, then we would expect the same scaling relation ($\dot{M}$ on $\tau_s$) to hold for $k_{\rm mod}$ when normalizing the measured accretion rates by \eqref{eq:kmod}. 

However, there is no proof that such a generalization is valid, though it has been implicitly assumed to hold in the pebble accretion prescriptions in global models of planet formation (e.g., \citealp{Morbidelli15}). The main caveat is that our normalization prescription is obtained essentially by assuming the particle motion is restricted at constant height, while in reality, particles follow complex trajectories as a result of turbulence. One of the main goals of this paper is to test whether pebble accretion efficiency remains the same in the turbulent 3D case as in the laminar 2D case.

In addition, given the qualitative nature of the derivation, the scaling relations are not necessarily very accurate, particularly near the transition between the drift and Hill regimes. More rigorous calculations can be found in \citet{Ormel2010}, who calculated the accretion rates via direct integration of individual particle orbits. Therefore, a more meaningful comparison is to directly compare $k_{\rm mod}$ among the hydro, AD, and ideal MHD runs.
 
\section{Simulation Results} \label{simresults}

\subsection{Accretion Rates} \label{Sec.acc_rate}

In Figure \ref{fig.k_coefficient_errbar_batch}, we show the steady-state accretion rate coefficients measured from our simulations (averaged over all particle injection cycles).
In our simulations with turbulence, individual cycles show significant temporal fluctuations in the pebble accretion rate. For the ideal MHD simulations in particular, no obvious steady state is achieved in individual cycles (this is also because of the limitation on the size of our simulation boxes, so that the time of continuous pebble flux is relatively short). With several cycles in each run, we expect that the standard deviation among the cycles approximately reflects the statistical uncertainties in our measurement of the pebble accretion rate. These are the uncertainties that we quote in Figure \ref{fig.k_coefficient_errbar_batch}. We may underestimate the uncertainties for the pebble accretion rate because of the uncertainty within each of the cycles, e.g., the fluctuation of accretion rate within the steady-state accretion stage in each cycle.

The absolute accretion coefficients shown in Figures \ref{fig.kabs_3_errbar_batch} and \ref{fig.kabs_2_errbar_batch} clearly indicate that there is a modest reduction of accretion rate in turbulence runs toward strongly coupled particles. Stronger turbulence leads to greater reduction. However, when looking at modified accretion rate coefficients shown in Figures \ref{fig.kmod_3_errbar_batch} and \ref{fig.kmod_2_errbar_batch},
the reduction is mostly eliminated. This fact suggests that turbulence does not strongly affect the {\it intrinsic efficiency} of pebble accretion. The reduction in absolute accretion rate with stronger turbulence is mainly caused by larger particle scale height. Other effects cancel out, as we discuss in Section \ref{sec-further}.
This is the most important conclusion from this work.

\begin{figure*}
  \centering

  \subfigure[AD simulation, core mass $\mu_c = 3\times 10^{-2}$.]
 {\label{fig.pass_nacc_AD2tau1}
    \includegraphics[height = 0.35\textheight]{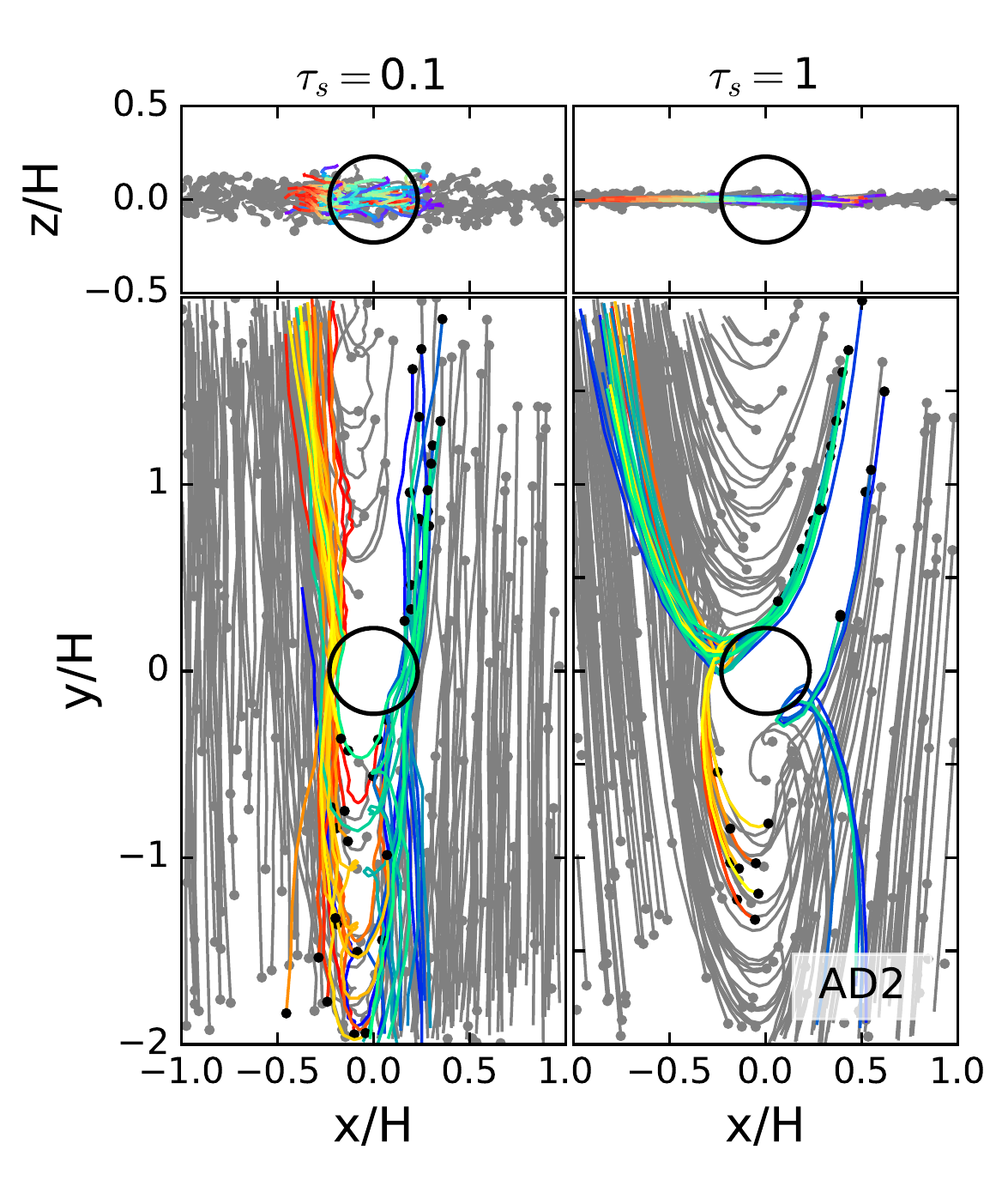}  
 }
  \subfigure[Ideal MHD simulation, core mass $\mu_c = 3\times 10^{-3}$.]
 {\label{fig.pass_idl3}
    \includegraphics[height = 0.35\textheight]{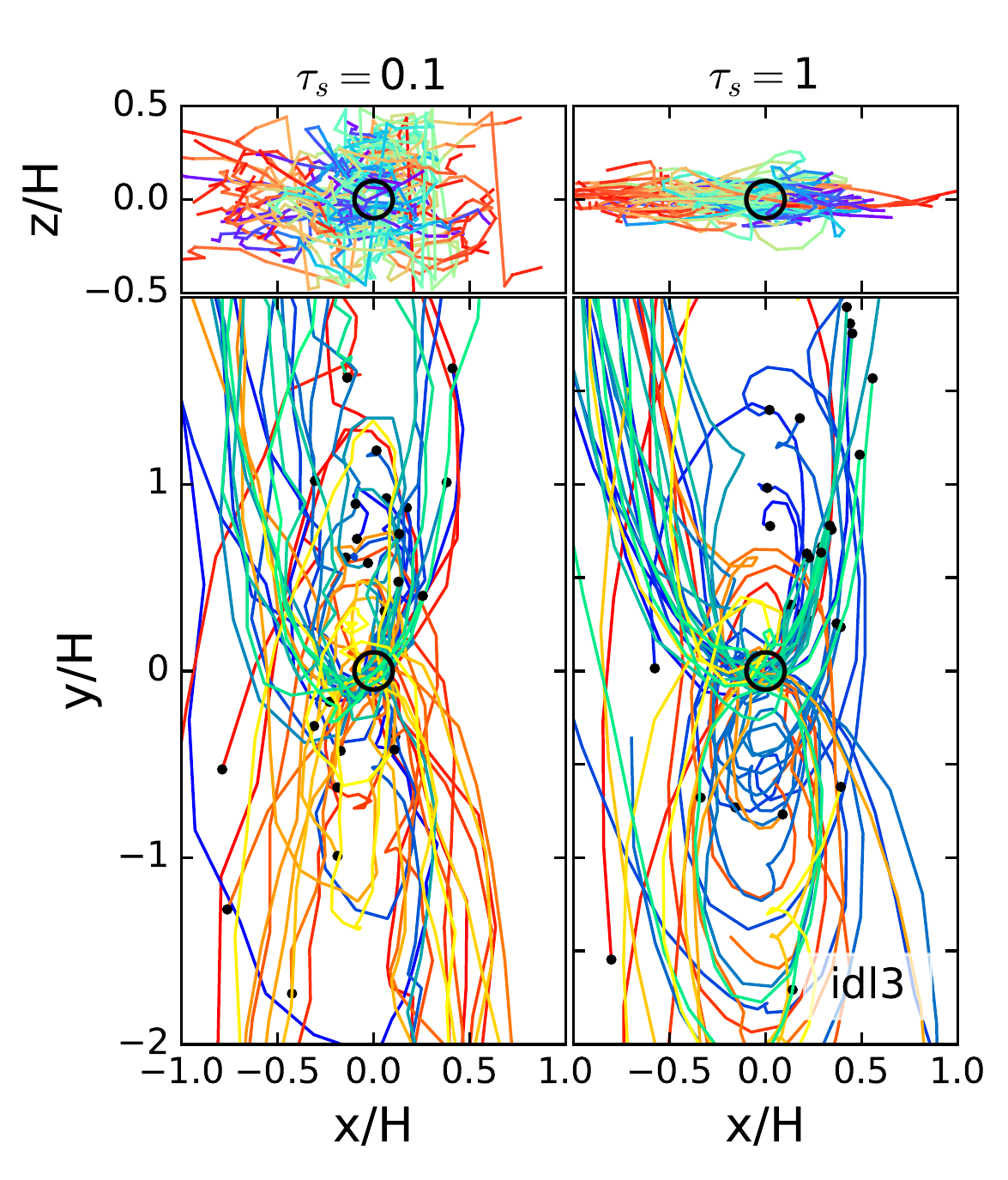}  
 }

 \caption{ \label{Fig.Traj_pass}Representative trajectories of particles that are not captured by the core in AD2 and idl3 simulations for particle sizes of $\tau_s = 0.1$ and $\tau_s = 1$. The black circles at the center of each panel mark the boundary of the Hill sphere. Colored trajectories in the x-y plane (lower panels) correspond to particles that enter the Hill sphere but are not accreted onto the core, and they are color-coded by the particles' initial positions (marked by black dots). Trajectories in the x-z plane (upper panels) are shown as cold to warm colors according to time. Gray lines in part (a) correspond to trajectories that do not enter the Hill sphere.}
\end{figure*}

Overall, the efficiency of pebble accretion peaks at $0.3\lesssim\tau_s\lesssim3$, consistent with theoretical expectations.
We also note from Figures \ref{fig.kmod_3_errbar_batch} and \ref{fig.kmod_2_errbar_batch} that the dependence of $k_{\rm mod}$ on $\tau_s$ at $\tau<1$ in pure hydrodynamic runs follows more closely to a power law of $\tau_s^{1/2}$, instead of $\tau_s^{2/3}$.
The main reason for the apparent discrepancy is likely because small particles with $\tau\lesssim0.3$ (0.03) in the $M_c=3M_t$ ($30M_t$) case are already in the drift regime of pebble accretion. Therefore, our simulations only probe a relatively narrow range in $\tau_s$ where the scaling relations are potentially applicable, and the difference between $\tau_s^{1/2}$ and $\tau_s^{2/3}$ is not very easily distinguishable. In addition, as discussed in Section \ref{ssec:theory}, the analytical scaling relations are unlikely to be accurate near the transition from the drift regime to the Hill one. 

With relatively small core mass $M_c=3\times10^{-3}M_T=3M_t$ (near the lower mass end of the Hill regime), we see that $k_{\rm mod}$ is slightly reduced toward smaller $\tau_s$ in runs with turbulence compared with the pure hydrodynamic run. Therefore, turbulence still leads to a small-to-modest reduction of the pebble accretion efficiency toward more strongly coupled particles (by a factor of at most 2-3 for $\tau_s\gtrsim0.01$).
With larger core mass (i.e., $M_c\sim30M_t$), on the other hand, we see that $k_{\rm mod}$ is insensitive to the strength of the MRI turbulence at least down to $\tau_s=0.01$. In other words, stronger gravity from more massive cores helps overwhelm the effect of turbulence, as one would naturally expect.

It is well known that pebble accretion is the most efficient for marginally coupled particles with $0.1\lesssim\tau\lesssim1$ in the Hill regime. We find that in some cases MRI turbulence even slightly enhances the accretion rate of particles at this size range. The reason will be discussed in the following subsections with more detailed analysis.

\subsection{Further Analysis}\label{sec-further}

The fact that the efficiency of pebble accretion is largely unaffected by turbulence is somewhat surprising, especially given that particles follow complex trajectories in the 3D case. In this subsection, we analyze the trajectories of accreted and non-accreted particles and address how turbulence affect the process of pebble accretion.

\subsubsection{Particle Trajectories}

In Figure \ref{Fig.Traj}, we show some typical trajectories of accreted particles with $\tau_s=0.1$ and $1$. Particles in each panel are randomly chosen to present the trajectories, so that the numbers of different ``types'' of lines shown in the figure roughly reflect the actual probability of each ``type'' of trajectory being populated. The Hill sphere is marked by a black circle in each panel. Projected to the x-y plane (lower panels in each subfigure), particle trajectories are color-coded based on their initial positions $(x_0,y_0)$ marked by the black dots, with warm (cold) colors corresponding to $y_0<0$ ($y_0>0$). 
We note that while initial particle distributions are spatially uniform, in practice, the pebble accretion process itself would inhibit some particle trajectories coming from the $y_0<0$ region (which is one limitation of our local simulations). Nevertheless, we note from Figure \ref{Fig.Traj} that these particles represent a minority among accreted particles, and we confirm that their contribution to our measured pebble accretion rates is well below $\sim50\%$.
Projected to the x-z plane (upper panels in each subfigure), each trajectory is colored from cold (blue) to warm (red) according to time.

We first see that for pure hydrodynamic runs, because particle orbits are sub-Keplerian and undergo radial drift, most accreted particles approach the core from the upper right side in each panel ($x_0>0, y_0>0$) corresponding to the upwind direction, and follow a parabolic shape before reaching the core due to shear.
As core mass increases, on the other hand, particles from the lower side can be captured as well, which is a result of a larger Hill radius ($\Omega r_H>\Delta v_K$), and can be deduced by directly integrating individual particle orbits \citep{Ormel2010}.

\begin{figure*}
  \centering
 \subfigure[Core mass $\mu_c = 3\times 10^{-3}$, particle size $\tau_s=0.1$.]
 {\label{fig.pos0_m3tau0.1}
    \includegraphics[width = 0.48\textwidth]{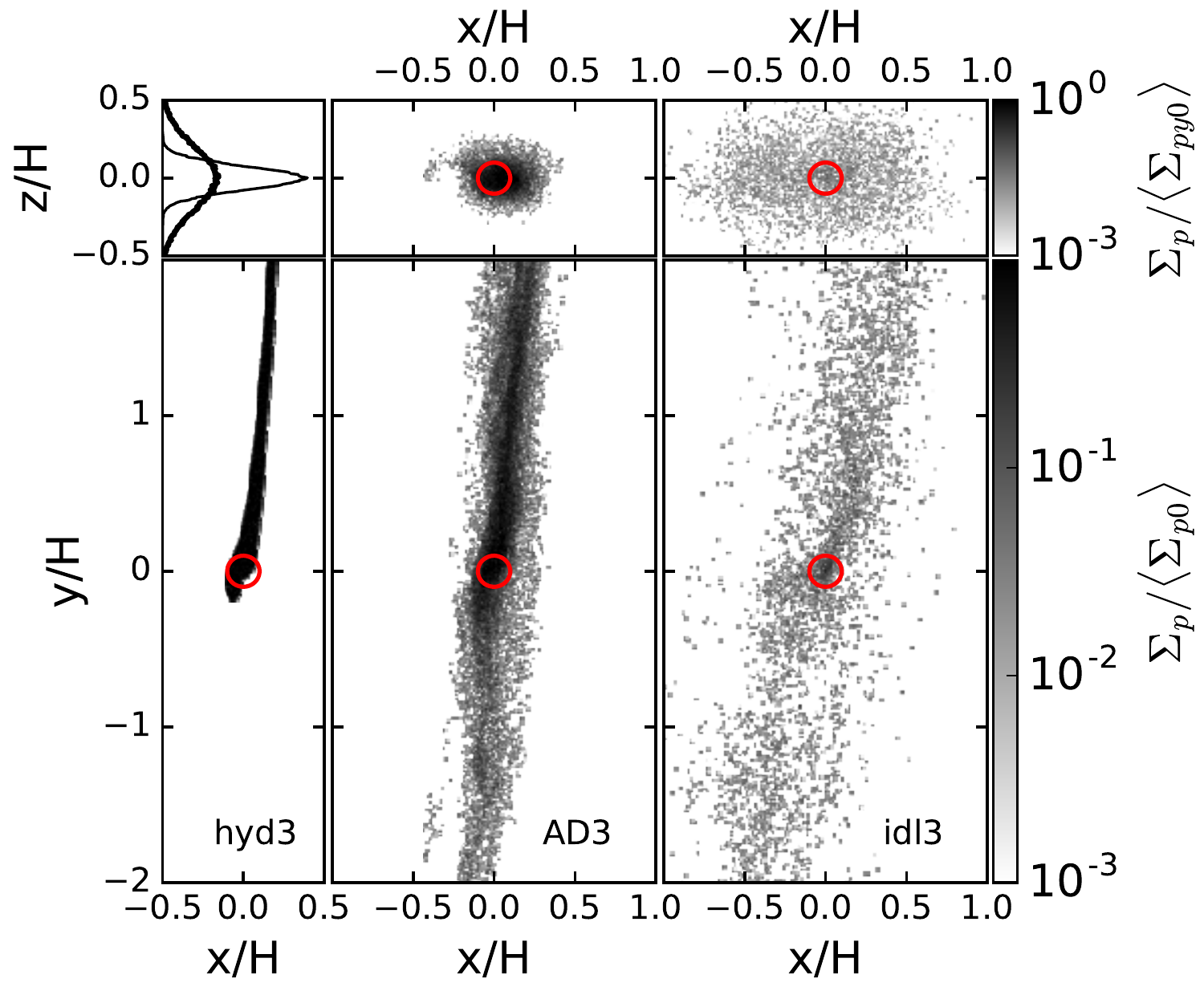}  
 }
 \subfigure[Core mass $\mu_c = 3\times 10^{-3}$, particle size $\tau_s=1$.]
 {\label{fig.pos0_m3tau1}
    \includegraphics[width = 0.48\textwidth]{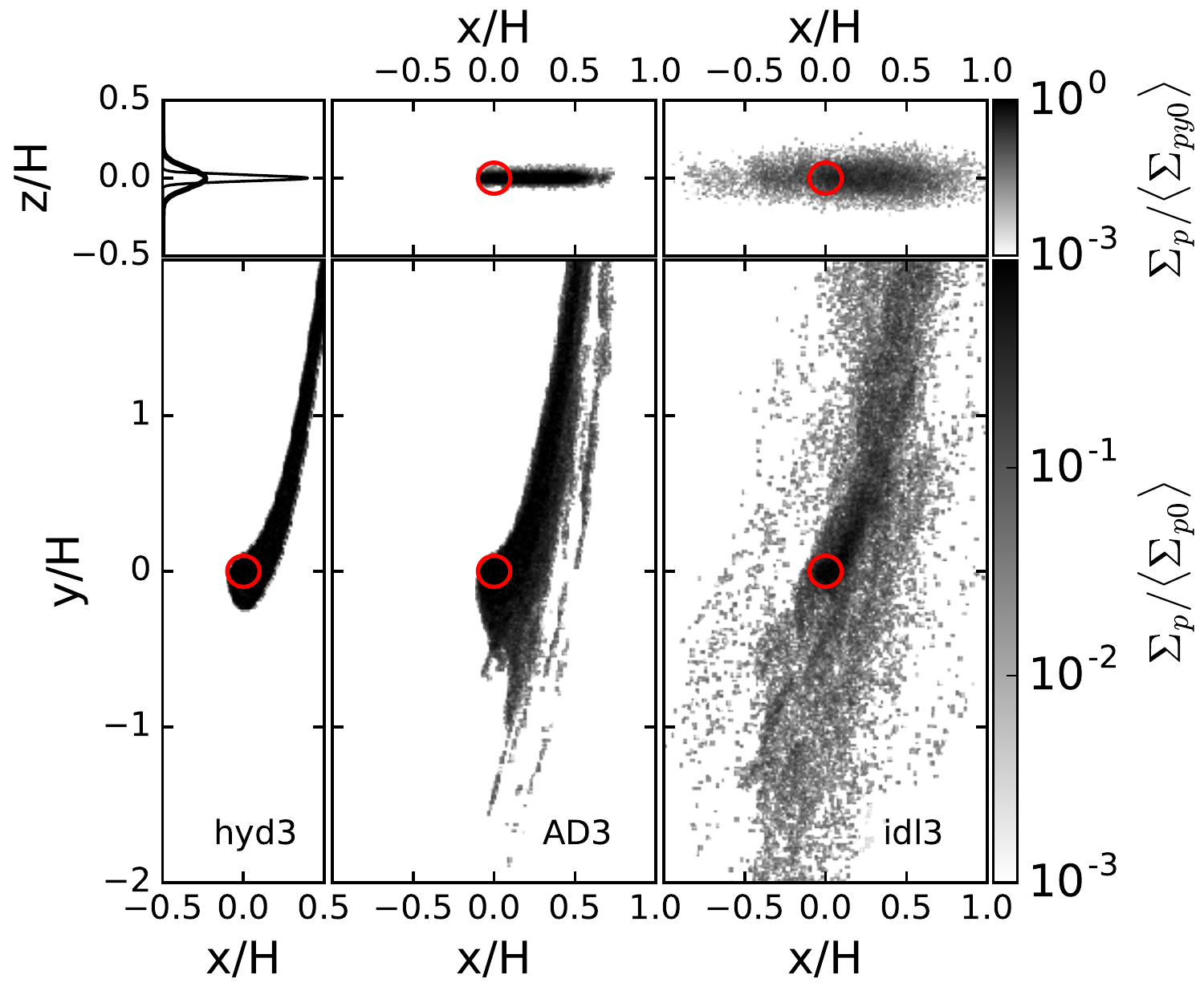}  
 }
 \subfigure[Core mass $\mu_c = 3\times 10^{-2}$, particle size $\tau_s=0.1$.]
 {\label{fig.pos0_m2tau0.1}
    \includegraphics[width = 0.48\textwidth]{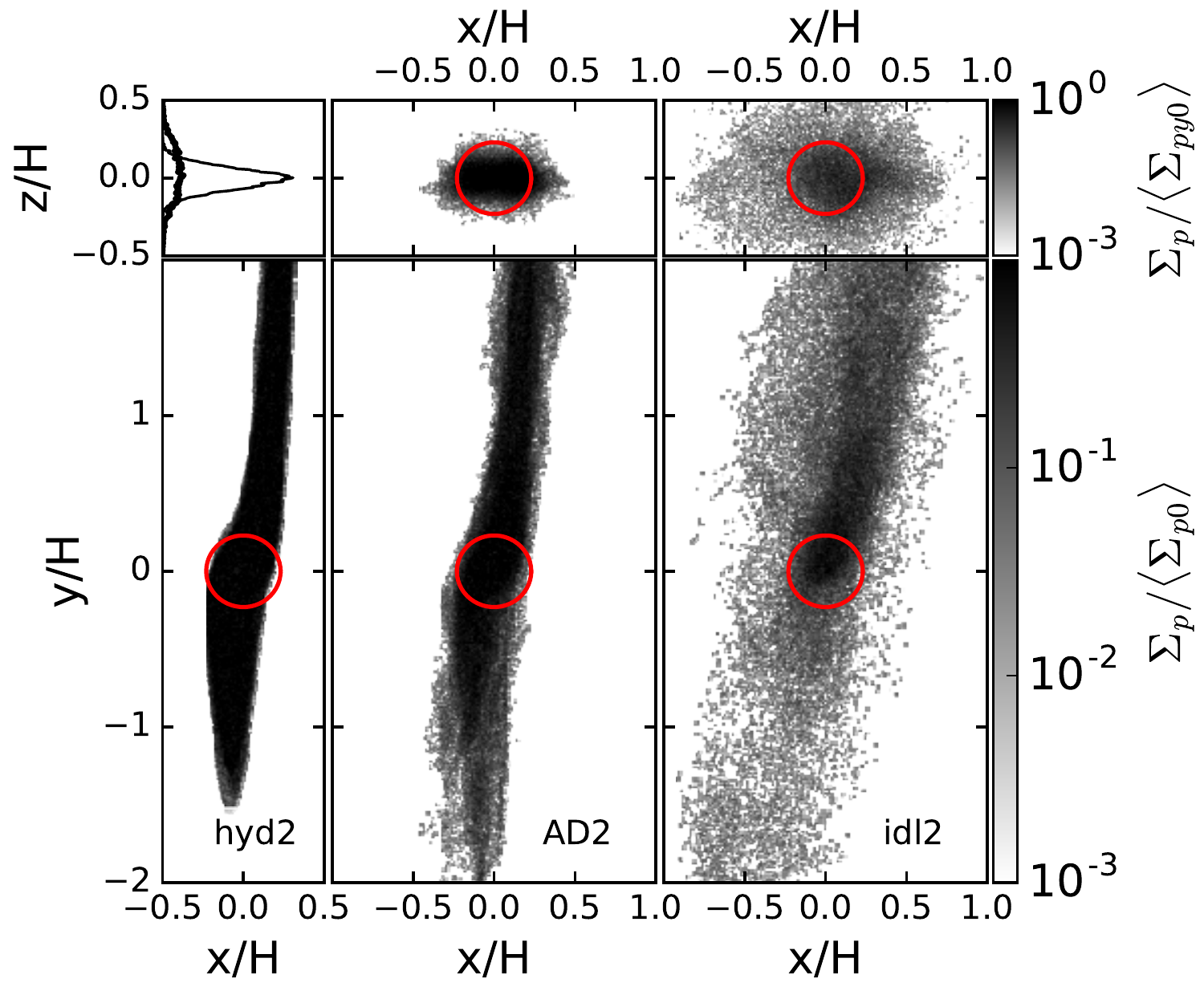}  
 }
 \subfigure[Core mass $\mu_c = 3\times 10^{-2}$, particle size $\tau_s=1$.]
 {\label{fig.pos0_m2tau1}
    \includegraphics[width = 0.48\textwidth]{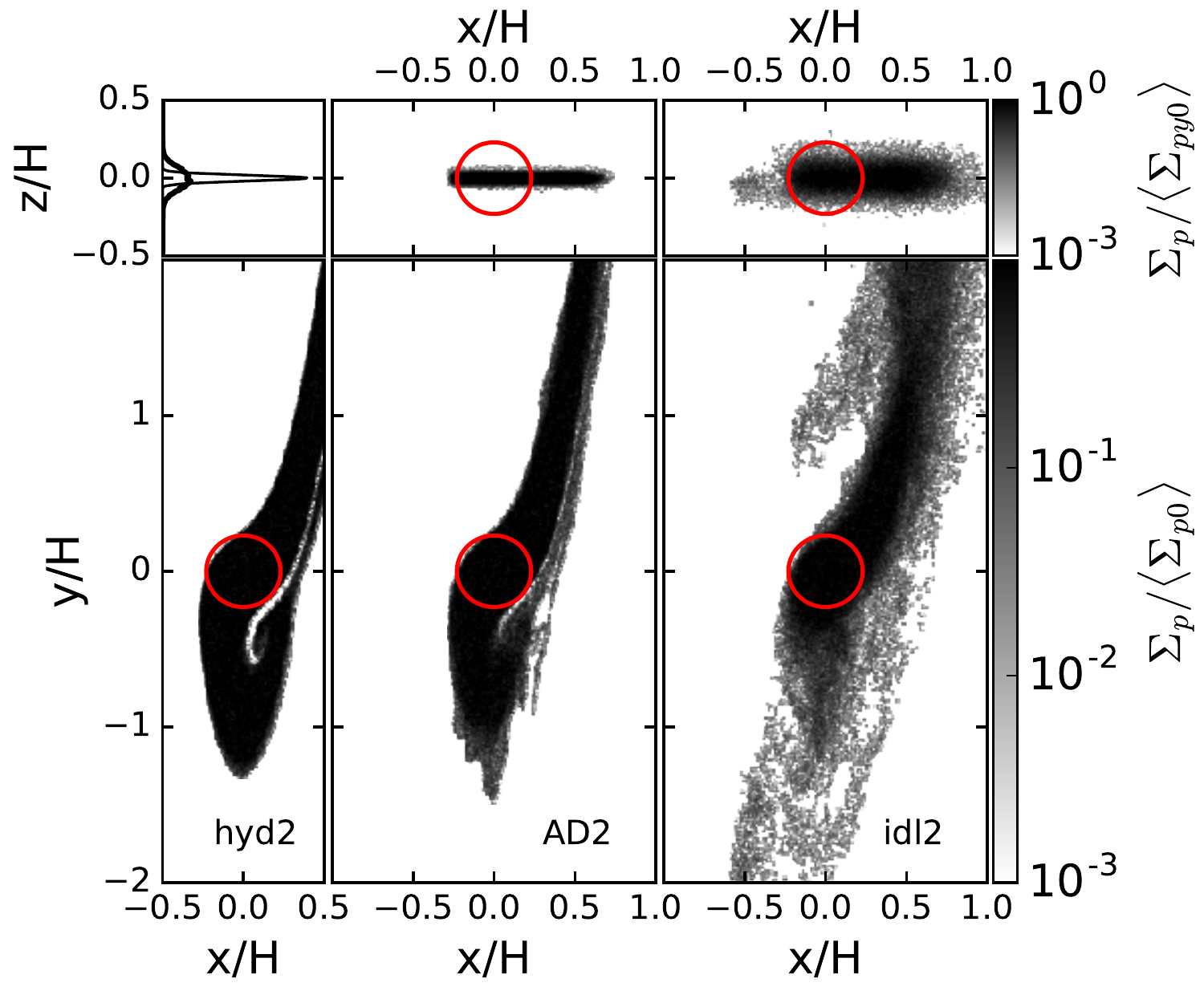}  
 }

 \caption{ \label{fig.pos0} Particle surface density (lower panels of each subplot) and azimuthally averaged density (two upper right panels of each subplot) of the $\tau_s=0.1$ and $\tau_s=1$ particles that are eventually accreted. The densities are normalized by the initial surface density or density profiles. The red circles at the center of the boxes indicate the boundary of the Hill sphere. The thick and thin curves in the upper left panel of each subplot show the initial vertical density profiles in the ideal MHD and AD simulations, respectively.}
\end{figure*}

Turbulence strongly affects particle trajectories in both the horizontal and in the vertical dimensions. In AD simulations with weak turbulence, while most trajectories are similar to those in the hydrodynamic run, we already see that the trajectories are more spatially spread, and more particles initially located in $y_0<0$ regions end up being accreted, implying that the zone feeding the planetary core becomes broader in the presence of turbulence. In ideal MHD simulations, particles show much more significant random motion, and their trajectories occupy a much larger volume of our simulation box (which requires us to choose a relatively large and broad box). Many particles exhibit swirling trajectories around $x\sim 0$ before reaching the core as they are scattered by turbulence. Overall, we see that stronger turbulence can bring particles from a broader range of locations to the core, which can potentially enhance the rate of pebble accretion.
  
On the other hand, not all particles that enter the Hill sphere can be accreted, especially in the presence of strong turbulence.
In Figure \ref{Fig.Traj_pass}, we show some typical trajectories of particles that have once entered the Hill sphere but eventually escaped capture by the core. We again choose particle stopping times $\tau_s=0.1$ and $1$, whose accretion radii $r_a$ are expected to be close to $r_H$. We show the trajectories only from runs AD2 and idl3 because they represent the two extreme situations with the most significant contrast. In run AD2, with the combination of a high-mass core and weak turbulence, we see that only a very small fraction of particles that enter the core escape, and these particles all barely touch the Hill sphere at the edge before exiting. By contrast, in run idl3, with the combination of a low-mass core and strong turbulence, we see that a large population of particles enter the Hill sphere from all directions. Their trajectories fill the entire Hill sphere yet they simply pass the core without being captured. Overall, we see that stronger turbulence reduces the probability for particles that enter the Hill sphere to be accreted.
  
In addition, we also see from Figures \ref{Fig.Traj} and \ref{Fig.Traj_pass} that when $H_p\gtrsim r_H$, particle trajectories experience significant vertical motion. 
This is the case for both accreted and non-accreted particles that enter the Hill sphere. The trajectories of $\tau_s=0.1$ and $1$ particles shown here span the entire vertical thickness of the particle layer. Therefore, even when accretion proceeds in a 3D manner, the core has potential access to a large fraction of particles in the entire particle column, instead of just those initially located within $|z|\lesssim r_H$ (as assumed in evaluating the normalization factor in Section \ref{kmod.meth}).

In sum, we have shown that turbulence affects particle trajectories in ways that can both enhance and reduce the rate of pebble accretion. The outcome of the pebble accretion efficiency is determined by the level to which these effects cancel. Enhancement or reduction of pebble accretion efficiency observed in Figure \ref{fig.kmod_3_errbar_batch} can be understood as imperfect cancellation, but overall, these effects cancel to a large degree and this leads to a more or less unchanged efficiency for pebble accretion in the presence of turbulence compared with the laminar case. 

\subsubsection{Map of Accretion Probability}

We can reinforce the conclusion reached in the previous subsection by looking at the map of accretion probability, shown in Figure \ref{fig.pos0} for particle sizes $\tau_s=0.1$ and $1$. The probability map is obtained by identifying the initial positions of particles that are accreted to the core, binning them to the grid, and normalizing them by the initial particle density. In practice, they are projected to the $x-y$ plane and the $x-z$ plane. 

For each core mass and particle size, the accretion probability in hydrodynamic simulations is either 0 or 1, with clear boundaries that indicate the shape of the feeding zones. Higher core mass leads to a wider feeding zone due to a larger Hill radius, and also enables the accretion of particles initially located on the downwind side of the flow, as mentioned earlier. The $\tau_s=1$ particles also have a slightly wider feeding zone than $\tau_s=0.1$ particles, which is consistent with theoretical expectations (see Section \ref{ssec:theory}). Besides, the feeding zones of $\tau_s=0.1$ particles are more aligned with the y-axis due to the particles' smaller radial drift velocities.

As the turbulence gets stronger, the map of accretion probability becomes smoother, and the feeding zone becomes broader. This is accompanied by the reduction in accretion probability within the feeding zone, especially for the ideal MHD case. Because our measurements show that $k_{mod}$ in the ideal MHD simulations is comparable with that in the hydrodynamic case, the two aforementioned effects approximately cancel each other. More specifically, with our fiducial core mass ($\mu=3\times10^{-3}$), 
we find that for $\tau_s=0.1$ particles, the fractions of particles entering the Hill sphere that are eventually accreted are $f_{hyd3}=0.77$, $f_{AD3} = 0.61$ and $f_{idl3} = 0.11$ in our hydrodynamic, AD and ideal MHD simulations, respectively; 
for $\tau_s=1$ particles, the corresponding fractions are $f_{hyd3}=0.91$, $f_{AD3} = 0.94$ and $f_{idl3} = 0.66$.
In the vertical direction, we see again that the core accretes particles initially located over most of the vertical column. 

Overall, our analysis of accretion probability supports our conclusion that the efficiency of pebble accretion is not strongly affected by turbulence because positive and negative effects that largely cancel each other.

\section{Summary and Discussion} \label{discussionandconclusion}

In this paper, we perform local unstratified shearing-box simulations to investigate
the effect of turbulence on the rate of pebble accretion. We focus on the Hill
regime, where pebble accretion is expected to be the most efficient, with planetary
core mass exceeding a transition mass (\ref{eq:Mtrans}). We consider turbulence
generated by the MRI either in the ideal MHD regime or with a non-ideal MHD effect
(AD), and compare the results with pure hydrodynamic simulations.
The primary goal of this study is to examine whether turbulence affects the
intrinsic efficiency of pebble accretion at the microphysical level, and to calibrate
analytical formulae on the rate of pebble accretion.

By initializing particles whose scale heights are predetermined from ``diffusion simulations", we conduct six simulations with different levels of turbulence and two different core masses. We choose core masses $M_c = 3\times 10^{-3}$ and $3\times 10^{-2}$ in units of the thermal mass, $M_T$, which are massive enough to accrete in the ``Hill regime" ($M_c > M_t$) but not massive enough to open a gap in the gas disk.  These values correspond to 0.5 and 5$M_\oplus$ for the minimum mass solar nebula. Our main results can be summarized as follows. 

\begin{itemize}
\item[1.] Overall, pebble accretion of marginally coupled particles ($\tau_s=0.1-1$) remains intrinsically efficient even under unrealistically strong MRI turbulence. 
\item[2.] The MRI turbulence reduces the efficiency of pebble accretion at a modest level toward strongly coupled particles ($\tau_s\lesssim0.03$)
and small core mass ($M_c$ just a few times $M_t$).
\item[3.]
For these core masses, the effect of  MRI turbulence on the  pebble accretion rate may be estimated by adjusting the particle scale height without having to calculate additional microphysics. The fact that the overall efficiency of pebble accretion is not strongly affected by the MRI turbulence is largely owing to a non-trivial cancellation of two effects resulting from the turbulence: an enhancement in the number of particles that are brought into the vicinity of the core, and a reduction in the probability that these particles get accreted.
\end{itemize}

Despite some uncertainties in our measured pebble accretion rates due to our limited simulation box size (leading to relatively brief period of steady-state accretion), the statistics is improved by applying multiple cycles of particle injection.

Our simulations have covered only a limited parameter space. The planetary core masses are chosen to be relatively large to allow pebble accretion to proceed mostly in the Hill regime. Moreover, we have fixed the sub-Keplerian velocity at $\Delta v_K=0.1c_s$, whereas real PPDs would encompass a wide range of values. 
While these choices are made to mimic conditions in the outer regions of PPDs, we are also limited by computational power: our simulations require both very high numerical resolution and a relatively large simulation box. In practice, high resolution is necessary only in the vicinity of the core's Hill sphere, and the use of a uniform grid in the Athena code is not optimal for our intended application. To extend our study and cover broader parameter spaces, for instance, for conditions toward smaller disk radii, it is highly desirable for future generations of the hybrid MHD-particle codes to have the capability of nested grids. With mesh refinement applied only to the vicinity of the core, it would be possible to more reliably determine the fate of more strongly coupled particles (for applications toward smaller disk radii, millimeter-centimeter sized pebbles already become strongly coupled with $\tau_s\lesssim0.01$). In particular, in the vicinity of embedded planetary cores, complex 3D flow structures have been found on the scale of the core's Bondi radius \citep{Ormel15,Fung2015}, which may affect the accretion of strongly coupled particles.

The application of pebble accretion theory to global models of planet formation (e.g., \citealp{Chambers14,Kretke2014,Morbidelli15}) has major uncertainties that lie in our lack of knowledge of disk substructure. In particular, the rate of pebble accretion depends sensitively on the radial pressure gradient (or $\Delta v_K$) in PPDs, and recent observations suggest that dust gaps and rings are common in PPDs (e.g., \citealp{ALMA2015,Andrews2016,Nomura_etal16,Zhang2016}), implying that the radial pressure gradient in disks is not smooth. Theoretically, it has been realized that disk structure and evolution are largely controlled by the amount and evolution of magnetic flux threading the disk (e.g., \citealp{Bai16}), and disk substructure may be associated with the phenomenon of magnetic flux concentration and zonal flows \citep{BaiStone2014,Bai15}. In the future, global disk simulations that incorporate relevant disk microphysics are essential to help establish a most realistic picture of PPDs, which will further provide essential input for global models of pebble accretion, and more generally for planet formation.
\\~\\
We thank Scott Kenyon for constructive conversations
and Anders Johansen for useful discussions.
X.N.B. acknowledges support from Institute for Advanced Study, Tsinghua University.
R.A.M.-C. acknowledges support from NSF grant number AST-1555385.

\bibliographystyle{apj}

\end{document}